\documentclass[a4paper,fleqn,usenatbib]{mnras}

\usepackage[T1]{fontenc}
\usepackage{ae,aecompl}

\usepackage{graphicx}	
\usepackage{amsmath}	
\usepackage{amssymb}	

\title[Stellar clusters in NGC 1316]{The complex star cluster system of NGC 1316 (Fornax A)}

\author[L. Sesto et al.]{
Leandro A. Sesto,$^{1,2,3}$\thanks{E-mail: sesto@fcaglp.unlp.edu.ar}
Favio R. Faifer$^{1,2,3}$
and Juan C. Forte$^{3,4}$
\\
$^{1}$Facultad de Cs. Astron\'omicas y Geof\'isicas, Univ. Nac. de La Plata, Argentina\\
$^{2}$Instituto de Astrof\'isica de La Plata, La Plata, Argentina\\
$^{3}$Consejo Nacional de Investigaciones Cient\'ificas y T\'ecnicas (CONICET), Argentina\\
$^{4}$Planetario `Galileo Galilei', Secretar\'ia de Cultura, Ciudad Aut\'onoma de Buenos Aires, Argentina
}
\date{Accepted XXX. Received YYY; in original form ZZZ}
\pubyear{2015}
\begin{document}
\label{firstpage}
\pagerange{\pageref{firstpage}--\pageref{lastpage}}
\maketitle

\begin{abstract}
This paper presents Gemini-$gri'$ high quality photometry for cluster
candidates in the field of $NGC~1316$ (Fornax A) as  part of a study that also includes GMOS spectroscopy. A preliminary
discussion of the photometric data indicates the presence of four stellar
cluster populations with distinctive features in terms of age, chemical
abundance and spatial distribution. Two of them seem to be the usually old 
(metal poor and metal rich) populations typically found in elliptical galaxies.
In turn, an intermediate-age (5 Gyr) globular cluster population is the dominant component
of the sample (as reported by previous papers). We also find a younger cluster 
population with a tentative age  of $\approx$ 1 Gyr.    
\end{abstract}
%
\begin{keywords}
galaxies: elliptical -- galaxies: star clusters -- galaxies: haloes

\end{keywords}
%
%
%
%
%
%

\section{Introduction}
\label{Intro}
%
 Once considered as `simple systems', Globular Clusters (GCs) are steadily leaving that
 characterization as more complex features of their stellar populations are discovered, e.g., 
  \citet{Carretta2015}. This situation emphasizes the problem of not only understanding their formation
 as individuals  but also in the context of galaxy formation \citep[e.g.][]{Brodie2014, Kruijssen2015, Kruijssen2016, Harris2015}.\\
 The idea that GC systems are connected with large scale  features of galaxies has its
 roots in \citet*{Eggen1962}. A recent example of this
 kind of analysis can be found in \citet{Forbes2016}.\\
 If GCs are in fact tracers of the dominant stellar populations formed in different events
 during the life of a galaxy, they should reflect some common features with field stars
 (e.g., in terms of ages, chemical abundances and spatial distributions).\\
 In this frame, $NGC~1316$,  a giant elliptical galaxy and strong radio source (Fornax A), appears
 as a particularly attractive object. On one side, the galaxy displays a number of morphological
 features that seem the fingerprints of `merger' activity (shells, ripples, complex dust lanes) 
 that have been studied in the optical range, for example, by \citet{Schweizer80, Schweizer81}.
 On the other, the galaxy exhibits a prominent GC system that has distinctive characteristics when compared with
 other bright ellipticals. The presence of `intermediate' age clusters in this galaxy, and their
 importance in the context of GCs formation, was already pointed out by \citet{Goudfrooij2001a}
 and subsequent studies \citep{Goudfrooij2001b, Goudfrooij2004, Goudfrooij2012}.\\
 A key feature in this analysis is the identification of the different kind of cluster systems that
 co-exist in $NGC~1316$. For example, \citet{Goudfrooij2001b} show that the integrated brightness-colour
 domain occupied by the `blue' GCs in this galaxy is very similar to that of the low metallicity and
 old halo clusters in the Milky Way (MW). In turn, that work also pointed out that `intermediate' colour
 GCs are considerably brighter in the average, then suggesting younger ages.\\
 Some of the features of the $NGC~1316$ GC system were studied by \citet{Gomez2001} on the
 basis of $BVI$ photometry. That work derived an overall ellipticity of 0.38 with a position
 angle of 63 degrees for the whole cluster system. Besides, their analysis of the projected areal
 density of the GCs  as a function of galactocentric radius suggests that both the `blue' and `red' 
 subpopulations share, to within the errors, very similar slopes.\\
 A more recent attempt to disentangle the GCs populations using wide-field $(C-R)$ photometry
 has been presented by \citet{Richtler2012b}. These authors explore the possible presence of
 two or three cluster populations and conclude that the dominant component is as young as
 2 Gyr.\\ 
 In turn, \citet{Richtler2014} presented a thorough study of the kinematic behaviour of the 
 GC system of $NGC~1316$ based on the radial velocities of 177 GCs.
 In another work, \citet{Richtler2012a} concentrated on the so called SH2 object, finding
 that it is in fact an unusual region of star formation.\\
 A recent estimate of the distance modulus of $NGC~1316$ has been presented by \citet{Cantiello2013}
 who derive $(m-M)_{0}=31.59$ (20.8 Mpc) by means of the SBF method, that we adopt in this paper, and 
 is somewhat smaller than that given in  \citet{Goudfrooij2001a}: $(m-M)_{0}$=31.80.\\ 
 In this work we present high quality Gemini $gri'$ photometry carried out on a CCD mosaic
 including eight different fields. This material is part of a study (in progress) that also includes
 GMOS spectroscopy for some 35 confirmed GCs. Currently, deep GC spectroscopic data is only available
 for three GCs \citep{Goudfrooij2001a}.\\
 Low photometry errors are crucial for a characterization of the different cluster populations. 
 In turn, the use of three different filters allows to account for field contamination and also for
 a comparison with simple stellar population (SSP) models in two colour diagrams.\\

 The paper is organized as follows: The characteristics of the data handling and photometry, including
 an analysis of the errors and completeness, is presented in Section \ref{sec2}. At the same time, in this section 
 we made the selection of unresolved sources and examined their colour-magnitude diagram.
 In Section \ref{sec3} we analyzed the distribution of the GCs on the sky as well as the colour-colour relations.
 An attempt to derive ages and chemical abundances using SSP models is presented in Section \ref{sec4}. The
 spatial distribution of each GC subpopulation and the  behaviour of the areal density profiles is discussed in 
 Section \ref{sec5}. A brief discussion of the radial velocities and of the GC  integrated luminosity function are given
 in Section \ref{sec6} and \ref{sec7}, respectively. Finally, a summary of the main conclusions is presented in Section \ref{sec8}.\\

\section{Data handling and photometry}
\label{sec2}
\subsection{Observations and Data Reduction}
\label{sec2.1}
\begin{figure}
	\includegraphics[width=\columnwidth]{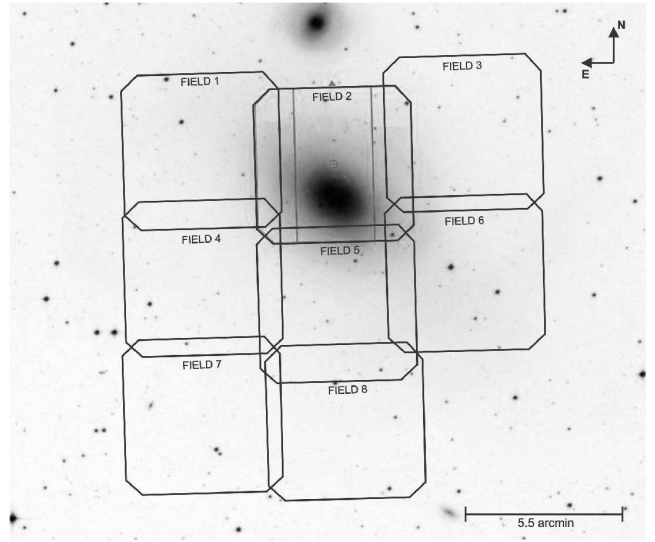}
    \caption{The positions of the eight GMOS fields superposed to the DSS image of $NGC~1316$.}
    \label{fig:fig1}
\end{figure}

The data set used in this work, listed in Table \ref{table_1}, was observed in image mode with the GMOS camera, mounted 
on Gemini South telescope, between September-October 2008 and August-October 2009
(Programs GS-2008B-Q-54 and  GS-2009B-Q-65, PI: J.C. Forte).\\
Four images per field with a binning of 2$\times$2 were taken through SDSS $g'$, $r'$ and $i'$ filters \citep*{Fukugita_1996}. These single exposures were obtained following a dithering pattern between them to facilitate cosmic-ray cleaning and to fill the gaps between the CCD chips. The distribution of the eight observed GMOS fields forms a mosaic as it is shown in Fig.~\ref{fig:fig1}.\\
\begin{table*}
\caption{Summary of observations. This table lists the identification number
of the fields, airmass, exposures and final seeing of the co-added images.}
\begin{tabular}{lcccccccccccccc}
\hline
\hline
\multicolumn{1}{c}{\textbf{Galaxy}} &
\multicolumn{1}{c}{} &
\multicolumn{1}{c}{\textbf{Field}} &
\multicolumn{1}{c}{} &
\multicolumn{3}{c}{\textbf{Airmass}} &
\multicolumn{1}{c}{} &
\multicolumn{3}{c}{\textbf{T$_{exp.}$(s)}} &
\multicolumn{1}{c}{} &
\multicolumn{3}{c}{\textbf{FWHM (arcsec)}} \\
&&&&$g'$&$r'$&$i'$&&$g'$&$r'$&$i'$&&$g'$&$r'$&$i'$\\
\hline
NGC1316&&1&&1.112&1.112 & 1.140&&4$\times$300&4$\times$150&4$\times$150&&0.90&0.83&0.80\\
&&2&&1.009&1.020&1.033&&4$\times$360&4$\times$200&4$\times$200&&0.85&0.75&0.67\\
&&3&&1.067&1.013&1.017&&4$\times$300&4$\times$150&4$\times$150&&0.75&0.72&0.63\\    
&&4&&1.090&1.050&1.067&&4$\times$300&4$\times$150&4$\times$150&&0.90&0.80&0.78\\
&&5&&1.180&1.024&1.035&&4$\times$360&4$\times$150&4$\times$150&&0.75&0.75&0.83\\
&&6&&1.435&1.319&1.255&&4$\times$360&4$\times$200&4$\times$200&&1.00&0.85&0.80\\
&&7&&1.111&1.069&1.048&&4$\times$360&4$\times$200&4$\times$200&&0.78&0.70&0.67\\
&&8&&1.072&1.115&1.150&&4$\times$360&4$\times$200&4$\times$200&&0.70&0.50&0.50\\
Standard&&E2-A&&1.043&1.044&1.045&&1$\times$10~~&1$\times$10~~&1$\times$10~~&&-&-&-\\
Blank sky&&-&&-&-&1.153&&-&-&7$\times$300&&-&-&-\\
\hline
\label{table_1}
\end{tabular}
\end{table*}
\begin{table*}
\caption{Multicolour photometry for all sources detected by SExtractor in the $NGC~1316$ fields. Magnitudes and colours are corrected for interstellar extinction. The complete photometry is available as Supporting Information with the on-line version of the article.}
\begin{tabular}{ccccccccc}
\hline
\hline
\textbf{ID}&\textbf{$\alpha_{J2000}$ ($h:m:s$)}&\textbf{$\delta_{J2000}$ ($\degr:\arcmin:\arcsec$)}&\textbf{$g'_{0}$}&\textbf{$Err g'_{0}$}&\textbf{$(g-r)'_{0}$}&\textbf{$Err (g-r)'_{0}$}&\textbf{$(r-i)'_{0}$}&\textbf{$Err (r-i)'_{0}$}\\
\hline
1&3:22:40.9&-37:13:43.8&24.313&0.029&0.535&0.043&0.459&0.046\\
2&3:22:48.9&-37:13:43.8&24.900&0.040&0.360&0.059&0.474&0.066\\
3&3:22:51.6&-37:13:43.8&25.483&0.050&0.276&0.088&0.309&0.129\\
\hline
\label{table_2}
\end{tabular}
\end{table*}
The reduction process includes the correction for instrumentals effects, the creation of mosaics  and the combination of different
 exposures for each filter. With that goal, the raw images were processed using the Gemini GMOS package within {\sc{iraf}}\footnote{IRAF is distributed by the National Optical Astronomical Observatories, which are operated by the Association of Universities for Research in Astronomy, Inc., under cooperative agreement with the National Science Foundation.} (e.g. {\sc{gprepare, gbias, giflat, gireduce, gmosaic}}), and applying the appropriate bias and flat-field corrections. The bias and flat-field images were acquired from the Gemini Science Archive (GSA) as part of the standard GMOS baseline calibrations.\\
The GMOS South EEV CCDs had significant fringing in the $i'$ filter. To subtract this pattern from our data, it was necessary
to create fringe calibration images from seven blank field images downloaded from the GSA (see Table \ref{table_1}). These baseline 
calibrations, although not taken on the same dates as those for $NGC~1316$, were sufficiently close to allow removing the
above mentioned effect. These frames were used to correct $i'$ images by means of the {\sc{gifringe}} and {\sc{girmfringe}} tasks.\\
Finally, in order to create a final image per field, suitable for photometry, all the individual exposures of the same filter
were co-added using the task {\sc{imcoadd}}.\\
\subsection{Photometry}
\label{sec2.2}
 As a previous step to the photometric measures, the luminous halo of $NGC~1316$ was removed using a script that combines features of SExtractor
 \citep{Bertin1996} and different tasks of {\sc{iraf}}, following the guidelines mentioned in \citet{Faifer2011}. This process also generates a list of  all sources detected by SExtractor.\\
Although all images used for the photometry showed high quality, the search for sources was conducted only on the $g'$ images, because the signal-to-noise ratio is slightly better than in the other bands (see subsection \ref{sec2.4}).\\
PSF photometry was performed using {\sc{daophot}}/{\sc{iraf}} routines \citep{Stetson1987}. The point spread function was determined through measurements of about two dozen of isolated and well-exposed objects located throughout the fields. In all cases, a Moffat25 model was adopted, since this model led to smaller rms than the Gaussian and Moffat15 options.\\
Once obtained the model that ensures the best fit, we run {\sc{allstar}} task to get PSF magnitudes for all objects detected by SExtractor.\\
As a final step, we used the {\sc{mkapfile}} task to derive suitable aperture correction for the PSF magnitudes.\\
\noindent
\subsection{Calibration}
\label{sec2.3}
The $NGC~1316$ photometry was transformed to the standard system using the E2-A standard star field from the `Southern Standard Stars for the $u'$ $g'$ $r'$ $i'$ $z'$  System' of Smith (2007)\footnote{http://www-star.fnal.gov/Southern\_ugriz/New/index.html}, which was observed on the same night as the central field (which contains the galaxy). These images were reduced using the same bias and flats-fields that were applied to our science data.\\
The standard field includes five stars with a $(g-r)'$ colour range from 0.5 to 1.05. \\
The photometric zero point calibration for all images observed under photometric conditions were derived adopting:

\begin{eqnarray}
\label{cero}
m_{std} = m_{zero} + m_{inst} - K_{CP} (X-1) + A_{p}
\end{eqnarray}\\

\noindent Where $m_{std}$ are the standard magnitudes, m$_{zero}$ is the photometric zero point, $m_{inst}$ are instrumental magnitudes,
 K$_{CP}$ is the mean atmospheric extinction at Cerro Pach\'on given by the Gemini web page\footnote{http://www.gemini.edu/?q=node/10445},
 $X$ is the airmass, and A$_{p}$ is the aperture correction for the PSF magnitudes.\\
 The analysis of the differences between the standard and instrumental magnitudes, so obtained, shows no significant trend with the colours of the standard stars in any of three $g'$, $r'$, $i$ bands, i.e., colour terms are zero (to within $\pm$ 0.02 mag) in agrement with previous results by the Gemini South staff.\\
For the remaining seven fields, we sought for sources in common between the central field. In each overlapped region we found between 10 and 15 isolated and well exposed objects. Once identified, we determined the offsets to bring them to standard photometric system obtained for the first one. The offsets applied were lower than 0.1 mag, except in filter $g'$ between the fields 2 and 3, where the correction was 0.35 mag. The $rms$ in all cases do not exceed 0.05 mag.\\
Subsequently, we  corrected  by interstellar extinction conside\-ring the value indicated by \citet{Schlafly2011}, A$_{g'}= 0.069$ mag, A$_{r'}=0.048$ mag, A$_{i'}=0.035$ mag, corresponding to a colour excess $E_{B-V}=$0.018.\\
The magnitudes and colours, corrected for interste\-llar extintion (denoted with the `0' subscript), for all sources detected by SExtractor in $NGC~1316$ are given in Table \ref{table_2}.\\
\begin{figure}
	\includegraphics[width=\columnwidth]{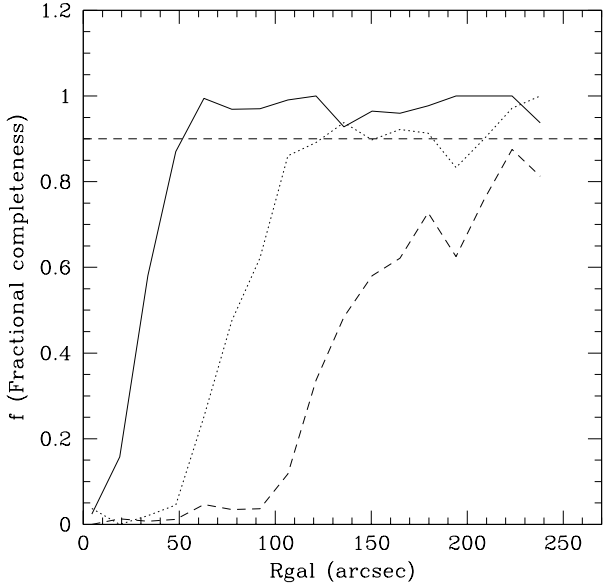}
    \caption{Completeness of the photometry as a function of the galactocentric radius
    for three different limiting magnitudes: $g'_{0}=23.5$ (solid line) $g'_{0}=25.0$ (dotted line) and $g'_{0}=25.8$ (dashed line). 
    The horizontal line indicates a 90 percent level.}
    \label{fig:fig2}
\end{figure}

\subsection{Completeness}
\label{sec2.4}

A completeness test was carried out in order to quantify the detection limits of our photometry.
Artificial stars were added to each $g'$ band image. They were distributed  with galactocentric
 radius following a power law (i.e. $r^{-1}$), in the fields that contain the galaxy. We selected this particular law, because it represents a good approximation of the slope that follows the GCs in their spatial distribution. In cases where the galaxy halo is weak, the artificial stars were added in a uniform way. For this process we used the {\sc{starlist}} and {\sc{addstar}} tasks.\\
The artificial objects were separated into bins of 0.1 mag covering a range in $g'_{0}$ from 18 to 26 mag. Two hundred point sources were added to the original $g'$ images in each magnitude bin. We then performed the search for sources in the same way as the original frames and thereby we got the fraction of recovered objects for each magnitude range.\\
As an example, Fig.~\ref{fig:fig2} shows the results of the completeness tests, as a function of 
 galactocentric radius, for different limiting magnitudes $g'_{0}$. At $g'_{0}=23.50$ our photometry is almost 
 complete outwards a galactocentric radius $R_{gal}=45$ arcsec while, increasing the limiting magnitude to $g'_{0}$ to 25.0, 
 yields a completeness factor close to 90 percent outwards $R_{gal}=90$ arcsecs.\\
Subsequently, the analysis was performed as a function of the colour of the artificial objects, in order to evaluate the presence of a colour bias in the completeness level. We have chosen five fixed colour values, i.e., $(g-i)'_{0} =$ 0.0, 0.4, 0.8, 1 and 1.2, which practically comprise the entire colour range of GC candidates (see section \ref{sec3.2}). Our experiments show that for colours from 0.4 to 1.2, the behaviours of the completeness is similar to that mentioned in the previous paragraph for the whole mosaic. However in the field which contains the galaxy, we have found that our photometry has a lower completeness level for objects bluer than $(g-i)'_{0}=$0.4. Specifically, for objects with $(g-i)'_{0}=$ 0.0, our sample is complete at 80 percent for $R_{gal}>90$ arcsec at $g'_{0}=24$ mag.  In the same region, for objects with $g'_{0}>24$ mag, the completeness decreases  for this bluest group reaching 50 percent at $g'_{0}=25$.\\
\begin{figure}
	\includegraphics[width=\columnwidth]{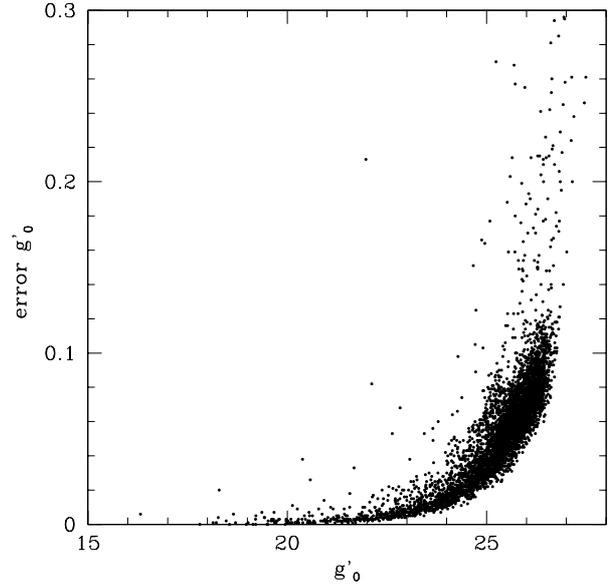}
    \caption{Photometry errors of the $g'_{0}$ magnitude for all the unresolved objects in the sample. At $g'_{0}=25.0$ the median error is $\approx$ $\pm$ 0.04 mag.}
    \label{fig:fig3}
\end{figure}

\begin{figure}
	\includegraphics[width=\columnwidth]{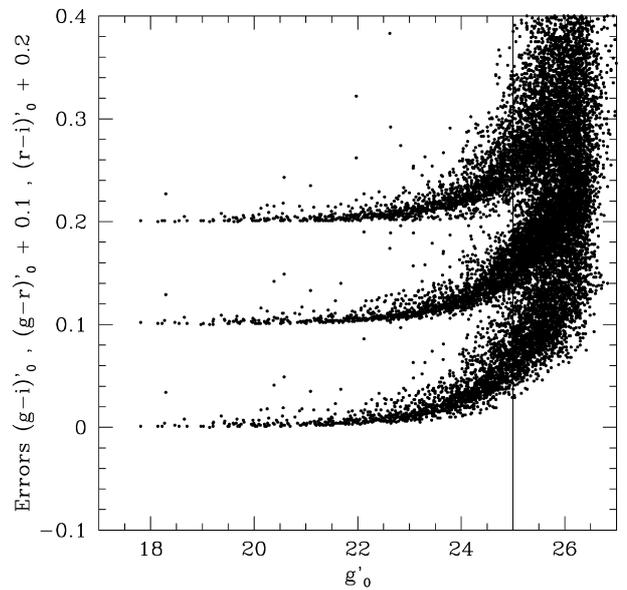}
    \caption{Photometry errors for the $(g-i)'_{0}$, $(g-r)'_{0}$ and
 $(r-i)'_{0}$ colour indices (arbitrarily shifted upwards) as a function of the $g'_{0}$ magnitude, for all the unresolved objects in the sample. The vertical line
 indicates the limiting magnitude of the analysis presented in this work.}
    \label{fig:fig4}
\end{figure}
\subsection{Selection of unresolved sources and the {$(g-i)'_{0} - g'_{0}$ Colour-Magnitude Diagram}}
\label{sec2.5}
\begin{figure}
	\includegraphics[width=\columnwidth]{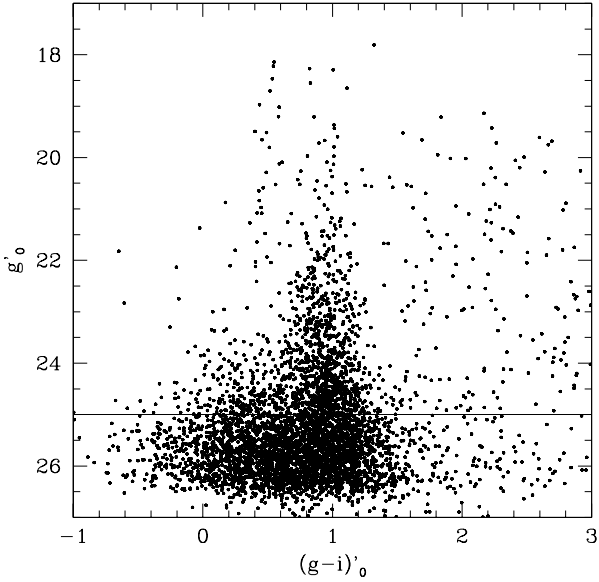}
    \caption{$g'_{0}$ vs $(g-i)'_{0}$ colour magnitude diagram for the unresolved sources (4856 objects). The horizontal line at $g'_{0}=25.0$ indicates the limiting magnitude of
     the analysis presented in this work.}
    \label{fig:fig5}
\end{figure}

At the distance of $NGC~1316$, we expect GC candidates to be unresolved sources. Therefore, the Stellarity index of SExtractor (0 for resolved objects and 1 for unresolved ones) was used to perform the object classification. We set resolved/unresolved boundary in 0.5.\\
In Fig.~\ref{fig:fig3} we show the error on $g'_{0}$ magnitudes for all the unresolved objects in the sample. The median error at $g'_{0}=25.0$ is $\approx$ $\pm$ 0.04 mag. In turn, the errors on the $(g-r)'_{0}$, $(r-i)'_{0}$ and $(g-i)'_{0}$ colours are displayed as a function of magnitude $g'_{0}$ in Fig.~\ref{fig:fig4}. The vertical line indicates the limiting magnitude of the analysis presented in this 
work, for which the median  photometric errors are $\approx$ 0.06 mag. Increasing the limiting magnitude leads to rapidly increasing errors in the colours, as shown in these last figures, conspiring against the detection of eventual structures in the GCs colour distribution.\\
The colour magnitude diagram, Fig.~\ref{fig:fig5}, corresponds to 4856 unresolved objects and exhibits some well known features already noticed in previous works, i.e., a broad colour distribution with an extended blue tail for GC candidates fainter than $g'_{0}$=23.5. Intermediate colour objects ($(g-i)'_0$ $\approx$ 0.90) include some GC candidates as bright as $g'_0$=19.0.\\ 
In particular, \citet{Goudfrooij2001b} find that an important fraction of the intermediate colour clusters is significantly brighter than their counterparts in the MW.\\

\section{Distribution on the sky and colour relations for GC candidates}
\label{sec3} 
%
\subsection{Distribution on the sky}
\label{sec3.1}
\begin{figure}
	\includegraphics[width=\columnwidth]{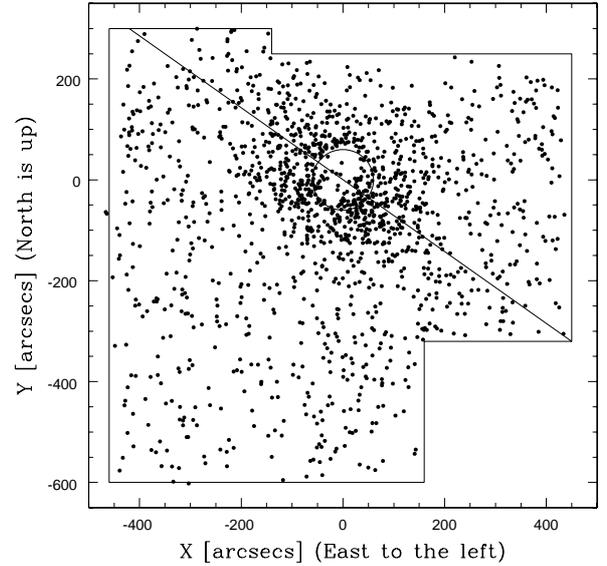}
    \caption{X and Y mosaic positions. Only `point' sources brighter than
     $g'_{0}= 25.0$ and $ 0.0 \textless (g-i)'_{0} \textless 1.35$ (1280 from a total of 4856 objects) are shown. The reference circle, centered on the $NGC~1316$
     nucleus ($\alpha_{J2000}$= 3$^{h}$22$^{m}$41.7$^{s}$; $\delta_{J2000}$= -37$\degr$12$\arcmin$28$\arcsec$), has a 60 arcsecs radius. The straight line indicates the position of the major axis of the galaxy ($PA=$55$\degr$).}
    \label{fig:fig6}
\end{figure}

The distribution of unresolved objects brighter than $g'_{0}=25$ (as well as the contour of the CCD mosaic) is 
displayed in Fig.~\ref{fig:fig6}. The reference circle is centered on the galaxy nucleus and has a radius of 60 arcsecs. The low
completeness within this region is a consequence of both the innermost dusty structure \citep[e.g.][]{Carlqvist2010, Asabere2014} and of the brightness of the nuclear region of the galaxy. The straight line with a position angle of 55 degrees,
corresponds to the position of the major axis of the galaxy derived with the  task {\sc{ellipse}} between galactocentric radii of 5 and 90 arcsec.\\
This figure shows a clear concentration around the galaxy centre and a detectable flattening of the spatial distribution of the point sources along the major axis of $NGC~1316$. This means that most of these sources in our sample are genuine GCs and that we are seeing the flattening already noticed by \citet{Gomez2001}.\\

\subsection{$(g-i)'_{0}$ Colour distribution}
\label{sec3.2}
\begin{figure}
	\includegraphics[width=\columnwidth]{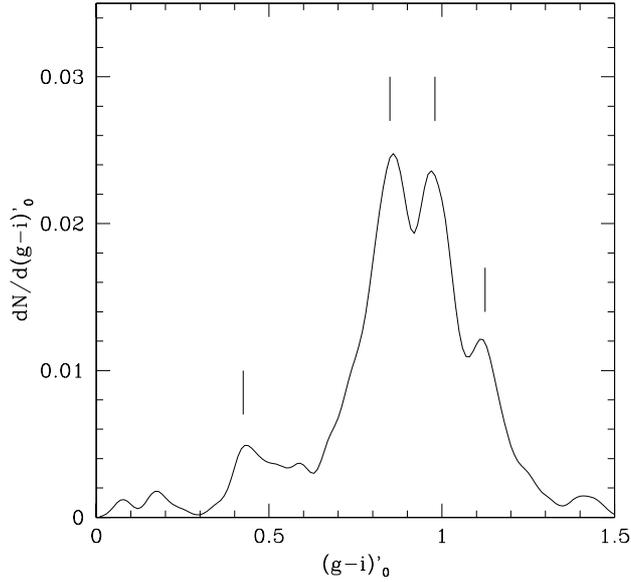}
    \caption{Smoothed $(g-i)'_{0}$ colour distribution for all objects in the mosaic
     field and brighter than $g'_{0}=23.50$. The four vertical lines at 0.42, 0.83,
     0.96 and 1.13, indicate the modal colours of the four stellar cluster populations
     discussed in the text.}
    \label{fig:fig7}
\end{figure}

\begin{figure}
	\includegraphics[width=\columnwidth]{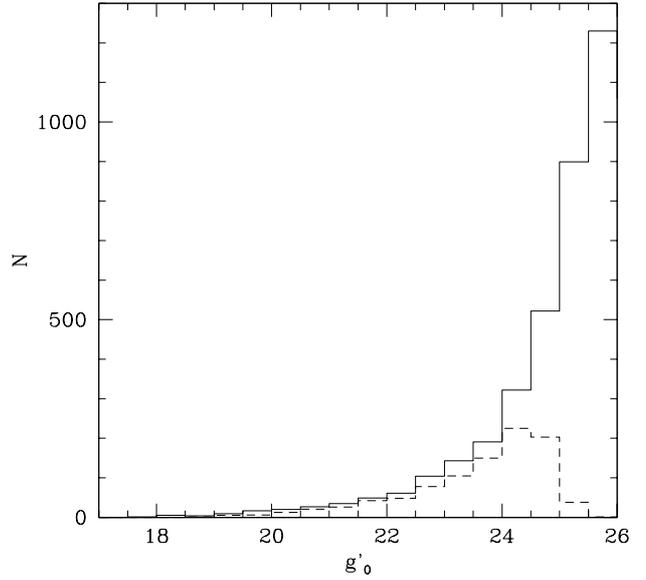}
    \caption{Comparison of the distribution of the $g'_{0}$ magnitudes presented in this work (solid lines)
  with that of the objects in common with \citet{Richtler2012b} (dashed lines).}
    \label{fig:fig8}
\end{figure}

\begin{figure}
	\includegraphics[width=\columnwidth]{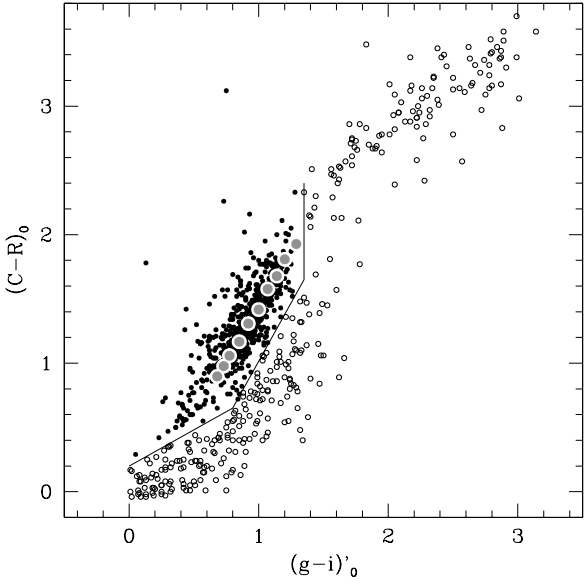}
    \caption{ $(C-R)_{0}$ vs $(g-i)'_{0}$ diagram for 990 objects in common with \citet{Richtler2012b}.
     Solid dots are GC candidates while open ones are considered as field objects. The broken line
     is a tentative boundary between both kind of objects. The gray dots display the $(C-T_{1})_{0}$ vs
     $(g-i)'_{0}$ relation determined from a GCs field in $NGC~4486$, shifted by -0.13 mag. in ordinates.}
    \label{fig:fig9}
\end{figure}

\begin{figure}
	\includegraphics[width=\columnwidth]{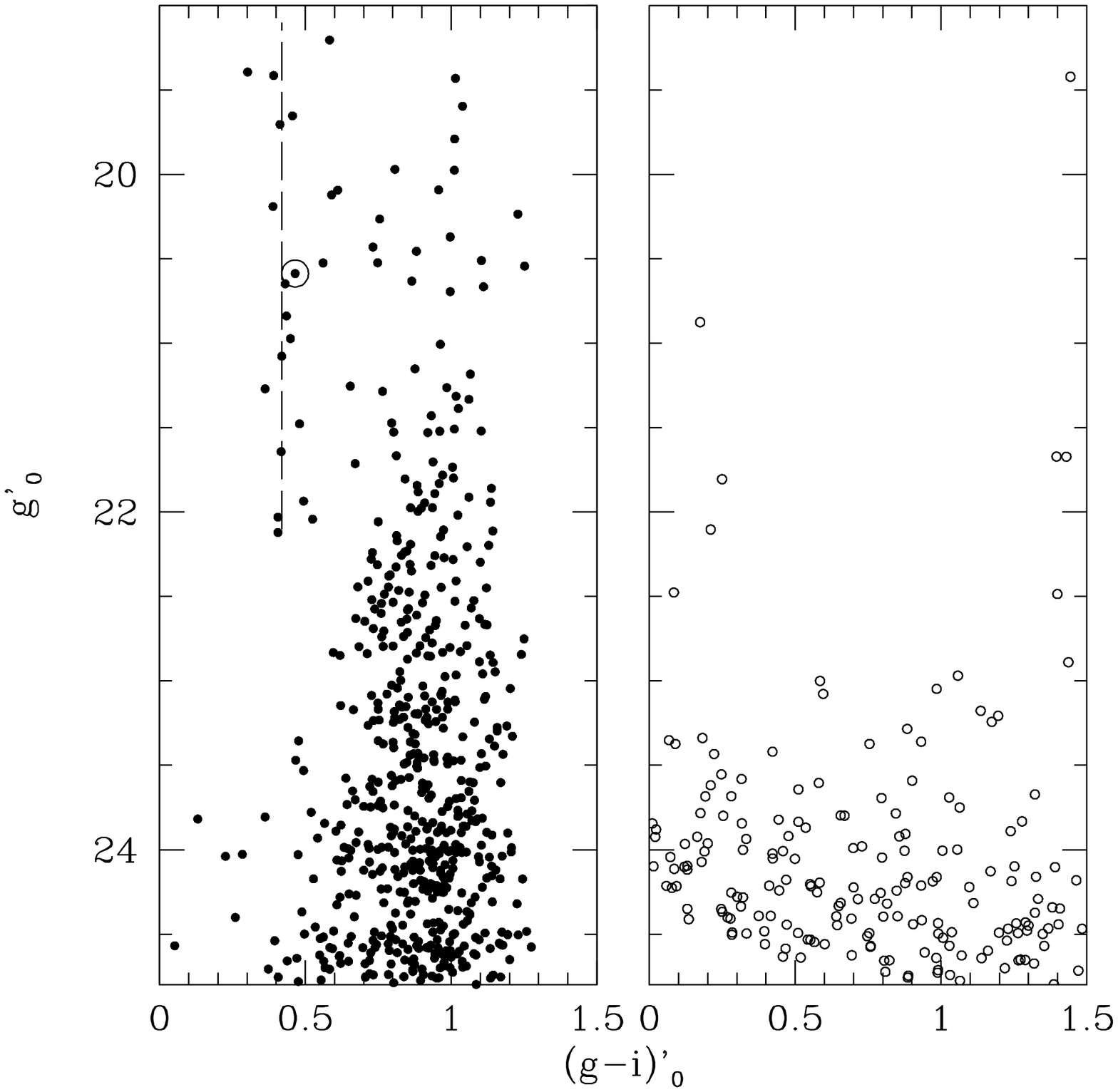}
    \caption{$g'_{0}$ vs. $(g-i)'_{0}$ colour magnitude diagrams for the objects shown in Fig. 9. 
    Left: GC candidates. The circled dot is the GC candidate number 119 in \citet{Goudfrooij2001a}. 
    Right: Field interlopers. The dashed line corresponds to the bluest peak in Fig. 7.}
    \label{fig:fig10}
\end{figure}

\begin{figure}
	\includegraphics[width=\columnwidth]{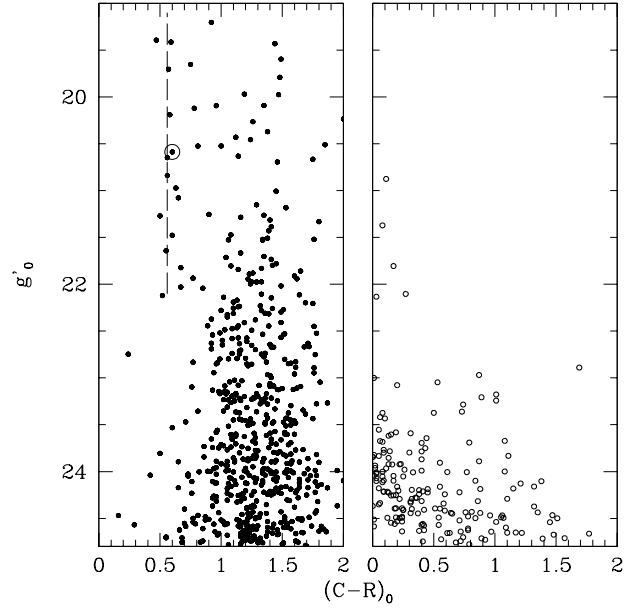}
    \caption{$g'_{0}$ vs. $(C-R)_{0}$ colour magnitude diagrams for the objects shown in Fig. 9.
    Left: GC candidates. The circled dot is the GC candidate 119 in \citet{Goudfrooij2001a}.
    Right: Field interlopers. The dashed line corresponds to the bluest peak in Figure 7.}
    \label{fig:fig11}
\end{figure}

\begin{figure}
	\includegraphics[width=\columnwidth]{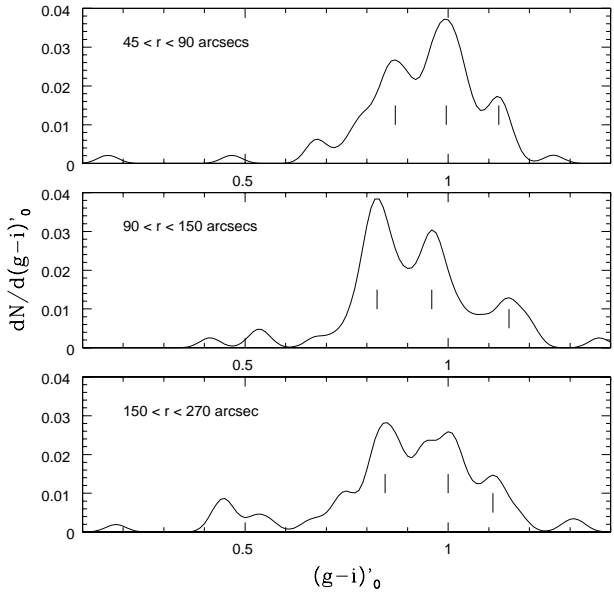}
    \caption{Smoothed $(g-i)'_{0}$ colour distribution for objects brighter than $g'_{0}=23.5$
     and galactocentric radius ranges of 45 to 90 arcsecs (upper panel), 90 to 150 
     arcsecs and 150 to 270 arcsecs. The vertical lines indicate the modal colour values.}
    \label{fig:fig12}
\end{figure}

\begin{figure}
	\includegraphics[width=\columnwidth]{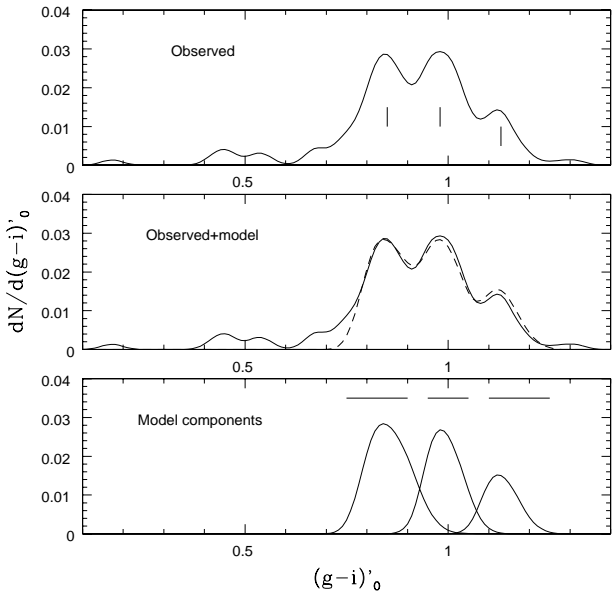}
    \caption{Normalized $(g-i)'_{0}$ colour distribution for all the objects included in the
    previous figure (upper panel). The vertical lines show the modal colours of the blue, intermediate
    and red GCs. The middle panel displays a comparison with the model fit described in the text. The three
    components of this model are shown in the lower panel.}
    \label{fig:fig13}
\end{figure}

\begin{figure}
	\includegraphics[width=\columnwidth]{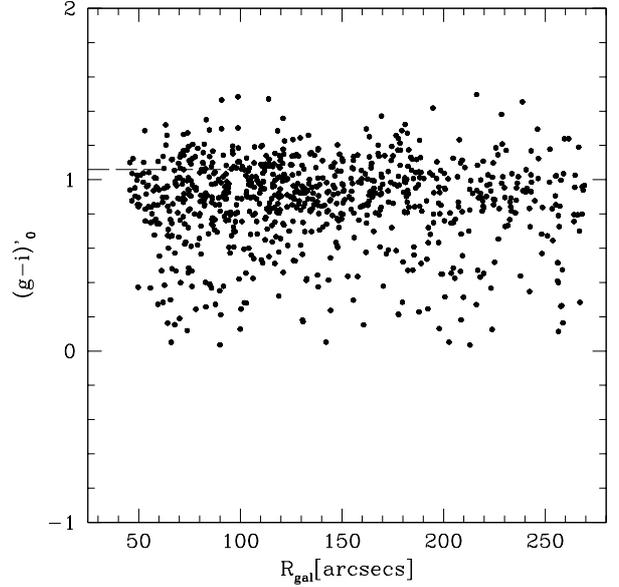}
    \caption{$(g-i)'_{0}$ colours as a function
     of galactocentric radius for GC candidates brighter than $g'_{0}=25.0$. This diagram
     shows that blue GC candidates are more abundant in the inner region of the galaxy.
     The dotted line indicates the colour of the galaxy halo within 100 arcsecs in galactocentric radius (see text).}
    \label{fig:fig14}
\end{figure}

\begin{figure}
	\includegraphics[width=\columnwidth]{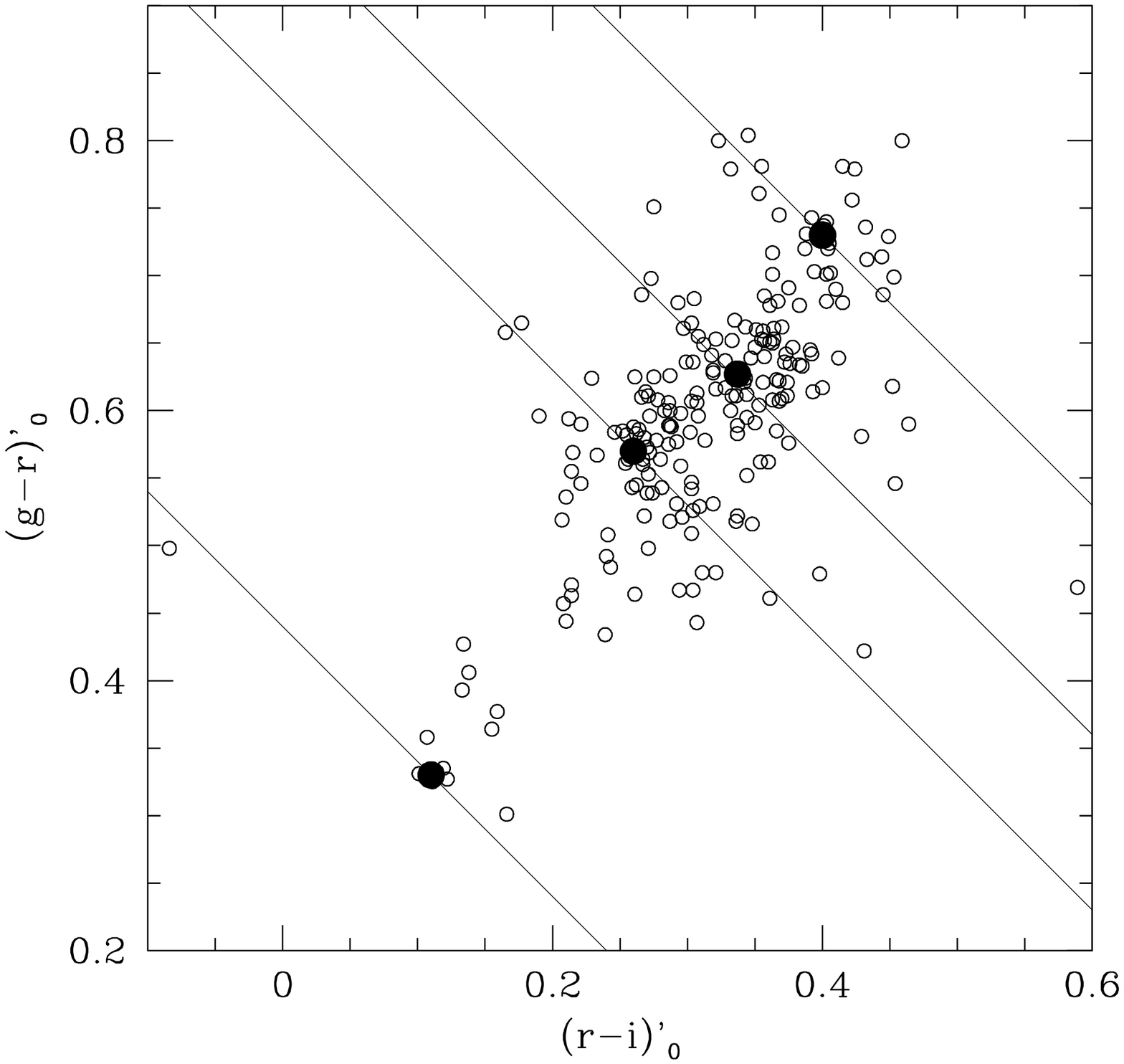}
    \caption{$(g-r)'_{0}$ vs $(r-i)'_{0}$ colours  for
    objects brighter than $g'_{0}=23.5$ and in the galactocentric range from 45 to 270
    arcsecs. The lines correspond to the modal $(g-i)'_{0}$ colours of the `very blue', `blue', `intermediate'
    and `red' cluster populations.}
    \label{fig:fig15}
\end{figure}

\begin{figure}
	\includegraphics[width=\columnwidth]{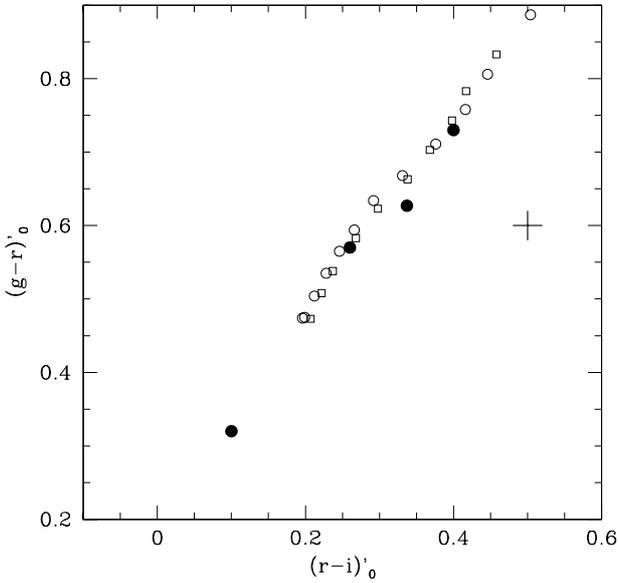}
    \caption{Modal $(g-r)'_{0}$ and $(r-i)'_{0}$ colours for the `very blue', `blue', `intermediate' and
     `red' cluster populations (filled dots). Open squares represent the modal colours
     determined for GCs in the field of $NGC~4486$ (M87). Open dots: PARSEC
     model 12 Gyr isochrone, after correcting $g'_{0}$ magnitudes (see text). The cross indicates typical
     errors for the observed modal colours.} 
    \label{fig:fig16}
\end{figure}

 Old GC systems in elliptical galaxies exhibit typical colour ranges $(g-r)'_{0}=$ 0.3 to 0.95,
 $(g-i)'_{0}$ 0.4 to 1.4 and $(r-i)'_{0}=$ from 0.0 to 0.6 \citep[e.g.][]{Faifer2011, Kartha2014, Harris2009}.  In the case of $NGC~1316$, where the
 existence of a young cluster population has been reported, and as a first approach, we extend
 the bluest limit to $(g-i)'_{0}=0.0$.\\
 In particular, the colour distribution of the GCs in $NGC~1316$ has been described as `unimodal'
 although `two colour peaks' can be detected for the brightest objects \citep[e.g.][]{Richtler2012b}.\\
 In addressing this issue, we first looked for a limiting magnitude $g'_{0}$ that could guarantee 
 both low photometric errors and a low contamination level by field interlopers.\\
 Fig.~\ref{fig:fig7}
 displays the smoothed $(g-i)'_{0}$ colour distribution for objects brighter than $g'_{0}=$ 23.5, adopting a
 Gaussian kernel of 0.025 mag, comparable to the photometric error of the $(g-i)'_{0}$ colour. This diagram shows
 three well defined peaks at 1.13, 0.96, 0.83 and a fourth, less evident one, at 0.42.\\
 In what follows, we attempt to compare these colour peaks with those found  by \citet{Richtler2012b} on
 the basis of their $(C-R)$ colours. Our photometry includes a total of 990
 objects in common with those authors.  The distribution of the 
 $g'_0$ magnitudes for these objects is displayed in Fig.~\ref{fig:fig8}. This diagram shows that both
 samples are very similar down to $g'_{0}=24.0$. At this magnitude, our errors on the $(g-i)'_{0}$
 colours are about half of those of the $(C-R)$ photometry. Besides, between $g'_{0}$ 24 and 25, our
 completeness is about two times larger than that of Richtler et al.\\
 The relation between the $(C-R)_{0}$ and $(g-i)'_{0}$ colours is displayed in Fig.~\ref{fig:fig9}. This diagram
 also includes the $(C-T_{1})_{0}$ vs $(g-i)'_{0}$ colour sequence determined by \citet{Forte2013}
 shifted by $-0.13$ mag. in ordinates. Most of the `genuine' GCs in $NGC~1316$ fall on this last sequence.\\
 From this diagram, we find that the color peaks determined by Richtler et al. at $(C-R)=$ 1.1 and 1.4
 correspond to those at $(g-i)'_{0}$ 0.83 and 0.96 in  Fig.~\ref{fig:fig7}.\\
 The broken line in Fig.~\ref{fig:fig9} is a tentative boundary between clusters and field objects. CG
 candidates and field objects, so defined, have clearly distinct behaviours in the colour magnitude diagrams displayed
 in Fig.~\ref{fig:fig10} and Fig.~\ref{fig:fig11}, where the $g'_{0}$ magnitudes are plotted vs $(C-R)_{0}$ and $(g-i)'_{0}$, respectively.
 The vertical lines centered at $(C-R)_{0}$=0.55 and $(g-i)'_{0}=$0.43 correspond to the bluest peak shown in  Fig.~\ref{fig:fig7}.\\
 The last two figures include a cluster (GC 119) with confirmed membership in $NGC~1316$ according to its radial
 velocity \citet{Goudfrooij2001a} that, as noted by \citet{Richtler2012b}, could be a very young cluster.\\
 As a first step, we analyze the global properties of the cluster system in
 an area defined by an outer galactocentric radius of 270 arcsecs, for which we have
a complete areal coverage and a completeness level that reaches $\approx$ 90 percent at
45 and 90 arcsecs for objects brighter than $g'_{0}$ = 23.50 and 25.00 mag, respectively.\\
The  $(g-i)'_{0}$ colour distribution of cluster candidates brighter than $g'_{0}=23.5$ mag within
annular regions defined between 45 to 90, 90 to 150 and 150 to 270 arcsecs are displayed in
Fig.~\ref{fig:fig12}. These distributions are normalized to the total number of clusters in
the $(g-i)'_{0}$ colour range from 0.00 to 1.35.\\
This diagram shows that the two more prominent peaks in the colour distribution for objects with 
galactocentric radius between 90 and 150 arcsec, are 0.03 mag bluer than those at the inner and
outer annuli. This feature is also observed when the limiting magnitude of the GC sample is increased
to $g'_{0}=25.0$, suggesting the presence of differential reddening possibly arising in the complex structure
of shells and ripples described by \citet{Schweizer81}. In what follows, however, and being a small
colour shift, we do not attempt any further correction to the $(g-i)'_{0}$ colours.\\
In turn, Fig.~\ref{fig:fig13}  shows the colour distribution of the whole GC sample between 45
and 270 arcsces (upper and middle panels).\\
 From the $(C-R)_{0}$ and $(g-i)'_{0}$ diagram, we estimate that, within the colour range of the
 GC candidates brighter than $g'_{0}=$ 23.50, the contamination by field interlopers is about 5 percent,
 and increases to 16 percent when that limit is increased to $g'_{0}=$25.0.\\
 The $(g-i)'_{0}$ colours of the GC candidates are displayed as function of galactocentric distance in Fig.~\ref{fig:fig14}. This diagram shows that
 the innermost regions of the galaxy exhibit a larger number of blue objects than the outer regions. This is compatible
 with \citet{Gomez2001} who found an `inverse' colour gradient (i.e. GCs become bluer inwards). A common feature in elliptical
 galaxies is  the existence of a rather constant lower blue boundary for the blue GCs while, for red GCs, the upper colour boundary becomes
 redder when the galactocentric decreases. This is opposite to the behavior displayed in Fig.~\ref{fig:fig14}. \\
According to the experiments mentioned in section \ref{sec2.4} the completeness level for objects bluer than $(g-i)'_{0}=0.4$ is lower than that for redder candidates. Therefore, this `inverse' colour gradient could be stronger than that seen in the figure.\\
 This figure also includes the colour of the galaxy halo inside 100 arcsec, $(g-i)'_{0}=$1.05, obtained from our
 innermost GMOS field. Due to the uncertainty in the sky brightness, this is just an indicative value and we are not 
 able to determine the eventual presence of a colour gradient, in fact detected by \citet{Richtler2012b} at galactocentric
 radii larger than 60 arcsecs.\\  
  Fig.~\ref{fig:fig15} displays the $(g-r)'_{0}$ vs $(r-i)'_{0}$ colours  corresponding to GC candidates
  brighter than $g'_{0}=23.5$ and galactocentric distances from 45 to 270 arcsecs. The straight lines in this 
  diagram indicate constant $(g-i)'_{0}$ colours corresponding to the four colour peaks shown in Fig.~\ref{fig:fig7}.\\
  In oder to determine the characteristic $(g-r)'_{0}$  and $(r-i)'_{0}$ colours of each of 
  the colour peaks, we looked for the modal colours along the (negative unit slope) straight lines, on which
  the $r'$ band colour  errors correlate, within bands of 0.05 mag wide and centered at the
  modal $(g-i)'_{0}$ colours.\\
  With this procedures we derive $(0.26 ; 0.57)$, $(0.33; 0.63)$ and $(0.40; 0.73)$ for the ($(r-i)'_{0}$; $(g-r)'_{0}$)
  colours of the `blue', `intermediate' and `red' clusters, respectively. These colours agree within $\pm$ 0.01
  mags with those derived when the limiting magnitude of the sample is increased to $g'_{0}=25.0$.\\
  In the case of the very blue clusters  we show an eyeball estimate:  $(g-r)'_{0}=0.32$; $(r-i)'_{0}=0.10$.\\  
  Monte Carlo simulations  indicate that the typical uncertainties of the modal colours, so determined, are 
  $\approx$ $\pm$ 0.015 mag.\\ 
  The four modal colours are shown in Fig.~\ref{fig:fig16} and compared with the revised colour-colour relation
  determined for GC candidates in a peripheral field of $NGC~4486$ by \citet{Forte2013}. The modal colours of the
 `blue' and `red' GC in  $NGC~1316$ fall close to that relation. In turn, the `intermediate' GCs seem shifted by -0.03
  mag in $(g-r)'_{0}$ and +0.04 in $(r-i)'_{0}$.\\
\section {Comparison with SSP models}
\label{sec4}
  This section presents a preliminary attempt to determine the characteristic ages
  and chemical abundances of the stella\-r cluster populations on a purely photometric
  basis.\\
  As well known, this kind of analysis suffers from the colour-age-chemical abundance
  degeneracy and any conclusion will require  a validation with the results
  from spectroscopic work. Furthermore, these results will be strongly dependent on the 
  adopted stellar population synthesis model.\\
\subsection{Model selection}
\label{sec4.1}
\begin{figure}
	\includegraphics[width=\columnwidth]{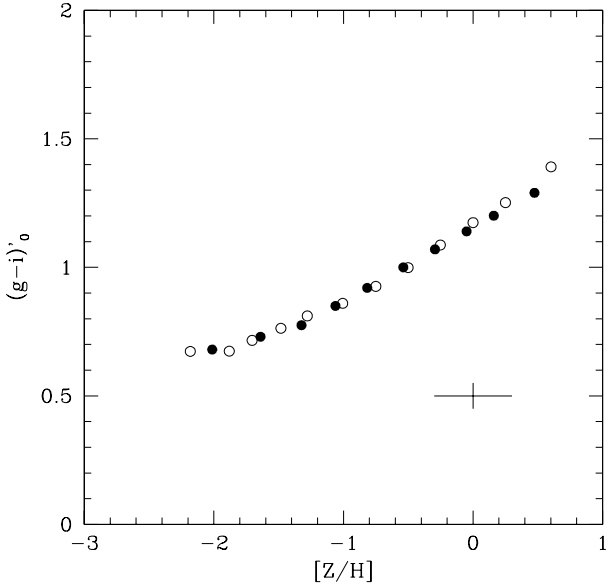}
    \caption{Chemical abundance $[Z/H]$ vs $(g-i)'_{0}$ colour relation based on the 
     \citet{Usher2012} calibration (filled dots) compared with the 12 Gyr PARSEC
     model including a correction to the $g'$ magnitude described in the text. The cross  
     shows typical errors for each data point.}
    \label{fig:fig17}
\end{figure}

\begin{figure}
	\includegraphics[width=\columnwidth]{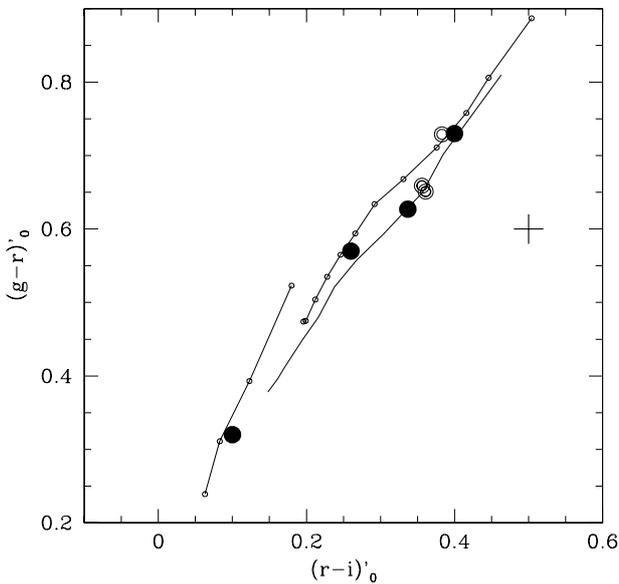}
    \caption{Corrected PARSEC model isochrone for 12 Gyr (upper line), 5 (lower line) and 1 Gyr (lower
    left line). The three large solid dots towards the upper right belong to the modal colours of the `blue',
    `intermediate' and `red' GC
    populations. The dot at $(r-i)'_{0}=0.10$, $(g-r)'_{0}=0.32$ is representative for the `very blue'
    clusters. Double circles are confirmed GCs with spectroscopic ages and chemical abundances
    from \citet{Goudfrooij2001a}. The cross shows typical errors of modal colours.} 
    \label{fig:fig18}
\end{figure}

  Several synthesis models for `single stellar populations' (SSP) are available in the literature
  and a thorough discussion of all their characteristics is beyond the scope of this paper.\\
  However, as a first step to choose a given model, we looked for those that provide
  a good representation of the several colour-colour relations observed in a  reference field in
 $NGC~4486$ \citep[e.g.][]{Forte2013}. The GC sample in that field is presumably dominated by old
 `bona-fide' clusters. \\
 The colour-colour relations presented in the last paper have been improved  to remove zero point differences
 between different works (Forte et al. in prep) and provide colour indices in the $SDSS$ 
 system. In particular, the updated $(g-r)'_{0}$-$(r-i)'_{0}$-$[Z/H]$ relation is given 
 in Table \ref{table_3}.\\
\begin{table}
\centering
\caption{Empirical colour-colour relation from GCs in $NGC~4486$. The chemical abundances $[Z/H]$
 are from the \citet{Usher2012} calibration.}
\begin{tabular}{c c c c c}
\hline
\hline
\textbf{(g-r)'$_{0}$}& &\textbf{(r-i)'$_{0}$}& &\textbf{[Z/H]}\\
\hline
0.473& &0.207& &-2.013\\
0.508& &0.222& &-1.640\\
0.538& &0.237& &-1.324\\
0.583& &0.268& &-1.062\\
0.623& &0.298& &-0.817\\
0.663& &0.338& &-0.538\\
0.703& &0.368& &-0.294\\
0.743& &0.398& &-0.050\\
0.783& &0.417& &0.160\\
0.833& &0.458& &0.474\\
\hline
\label{table_3}
\end{tabular}
\end{table}
 An overview of different SSP models shows that colour indices 
 observed for GCs in $NGC~4486$, and involving the $riz'$ bands, are fully consistent with the PARSEC
 models by \citet{Bressan12}  without requiring any zero point corrections.\\
 In turn, the PARSEC colour indices including the $g'$ band magnitudes, require a small correction that
 seems dependent on chemical abundance. In particular, we find that a correction to the model $g'$ magnitudes:\\
\begin{equation}
\Delta$g'= -0.04 ([Z/H]+2.2)$
\label{eq:correc}
\end{equation}
\\
\noindent to the 12 Gyr model with a Salpeter initial mass function, leads the model $(g-r)'_{0} - (r-i)'_{0}$ 
 relation to within $\pm$ 0.02 mag from the empirical relation in the whole colour range.\\
The maximum correction to the $g'_{0}$ magnitudes amounts to $-0.11$ mag in all range $[Z/H]$
(from -2.2 to 0.6) and its origin is not yet clear.\\
 The so corrected PARSEC models also show a small inflection in the $(g-r)'_{0} - (r-i)'_{0}$ plane, as
 shown later, that is not a common feature with other SSP models. This relation, for a 12 Gyr model, is given
 in Table \ref{table_4} and compared with the empirical one in Fig.~\ref{fig:fig16}.\\
Tables \ref{table_3} and \ref{table_4} also give the chemical abundances $[Z/H]$ based on spectroscopic 
observations of the Calcium triplet lines presented by \citet{Usher2012}. This `broken line' calibration, that has a change
of slope at $(g-i)'_{0}=0.77$, is in excellent agreement with the slightly curved relation predicted by the
PARSEC model, as displayed in Fig.~\ref{fig:fig17}.\\ 
\begin{table}
\centering
\caption{PARSEC model colour-colour-chemical abundance relation, for an age of
 12 Gyr and an assumed Salpeter initial mass function, and including a correction
 to $g'_{0}$ magnitudes (see text).}
\begin{tabular}{c c c c c}
\hline
\hline
\textbf{(g-r)'$_{0}$}& &\textbf{(r-i)'$_{0}$}& &\textbf{[Z/H]}\\
\hline
0.474& &0.196& &-2.182\\
0.475& &0.199& &-1.881\\
0.504& &0.212& &-1.705\\
0.535& &0.228& &-1.483\\
0.565& &0.246& &-1.279\\
0.594& &0.266& &-1.006\\
0.634& &0.292& &-0.750\\
0.668& &0.331& &-0.501\\
0.711& &0.376& &-0.252\\
0.758& &0.416& &0.000\\
0.806& &0.446& &0.250\\
0.887& &0.504& &0.602\\
\hline
\label{table_4}
\end{tabular}
\end{table}
 The four modal colours corresponding to the very blue, blue, intermediate and red GC
 populations are shown in Fig.~\ref{fig:fig18} together with the PARSEC models corresponding 
 to 1 Gyr (in the $[Z/H]$ range from -0.25 to 0.60) and the (corrected) 5 and 12 Gyr isochrones.
\subsection{Ages}
\label{sec4.2}
\begin{figure}
	\includegraphics[width=\columnwidth]{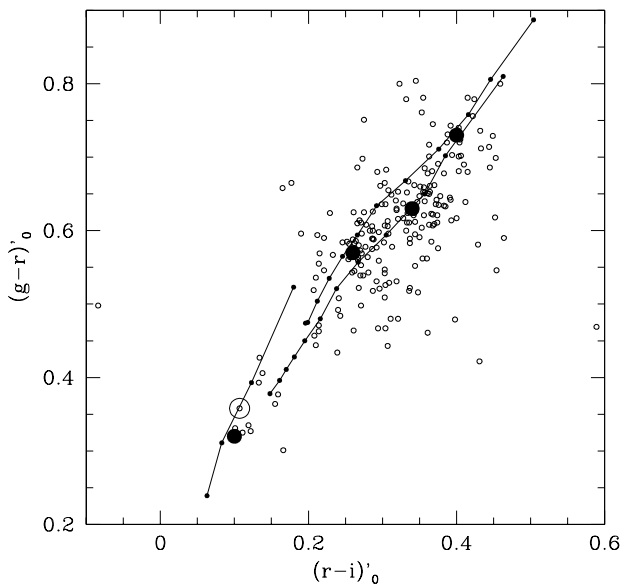}
    \caption{$(g-r)'_{0}$ vs $(r-i)'_{0}$ colours for 216 GC candidates brighter than $g'_{0}=$23.5 and 
    galactocentric radii beteween 45 and 270 arcsecs. The lines correspond (left to right) to the
    PARSEC models for 1, 12 and 5 Gyr. The encircled dot corresponds to the cluster candidate GC 119
    (see text).}
    \label{fig:fig19}
\end{figure}

\begin{figure}
	\includegraphics[width=\columnwidth]{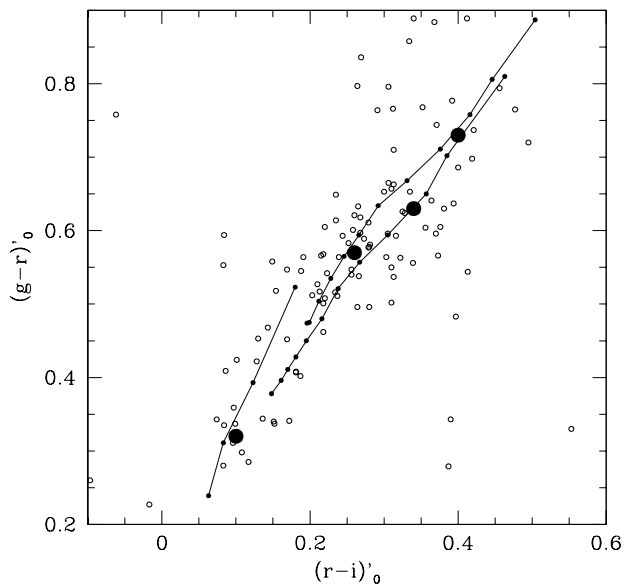}
    \caption{$(g-r)'_{0}$ vs $(r-i)'_{0}$ colours for 115 GC candidates brighter than $g'_{0}=$23.5 and 
    galactocentric radii larger than 270 arcsecs. The lines correspond (left to right) to the
    PARSEC models for 1, 12 and 5 Gyr.}
    \label{fig:fig20}
\end{figure}

 Each GC population has a characteristic age that, as previously said, cannot be unambiguously
 determined just from photometric data. At this stage, and until spectroscopic data is not available,
 we  make a `reasonable' (and possibly arguable) guess about the age of each GC subpopulation.\\
 In the case of the blue GC population, the colour peak at $(g-i)'_{0}=0.82$ is comparable to those observed
 in other early type galaxies studied with the same photometric system \citep[e.g.][]{Escudero2015, Kartha2014}.
 Besides, and as noticed by \citet{Goudfrooij2001a}, these clusters have photometric characteristics that are
 similar to those of the old blue GCs in the MW. With these arguments in mind, we adopt an age of 12 Gyr for
 the blue GCs in $NGC~1316$.\\
 For the intermediate GCs we already noticed that their modal integrated colours, shown in Fig.~\ref{fig:fig16}, are
 offset from those of the reference field in $NGC~4486$ and fall on the 5 Gyr ishochrone of the PARSEC models
 displayed in Fig.~\ref{fig:fig18}. This diagram also includes the only three intermediate GCs with spectroscopic
 age determinations, presented by \citet{Goudfrooij2001b}, for which they find and age of 3 Gyr. Two of these
 GCs are on the PARSEC 5 Gyr isochrone.\\
 Regarding red GCs, and as in the case of the blue GCs, we adopt a tentative age of 12 Gyr, a choice based
 on their brightness and colours, comparable to those observed in bright ellipticals.\\
 The same model isochrones are shown in Fig.~\ref{fig:fig19} and Fig.~\ref{fig:fig20} for GC candidates
 with galactocentric radii from 45 to 270 arcsecs and larger than 270 arcsecs, respectively. In these figures,
 the very blue cluster candidates appear close to the 1 Gyr isochrone. That isochrone displays $[Z/H]$ values of 
 -0.25, 0.0, 0.25 and 0.60 (blue to red) and, in particular, GC 119 appears near the colour of the very blue
 clusters peak and close the colours corresponding to solar metallicity. It must be stressed, however, that 
 the model ishochrone for an age of 1 Gyr is degenerate, i.e., lower metallicity models overlap in colour
 with those in the the range displayed in Fig.~\ref{fig:fig19} and Fig.~\ref{fig:fig20}.
 \subsection{Chemical abundances and the modelling of the blue, intermediate an red GC populations.}
\label{sec4.3}
\begin{figure}
	\includegraphics[width=\columnwidth]{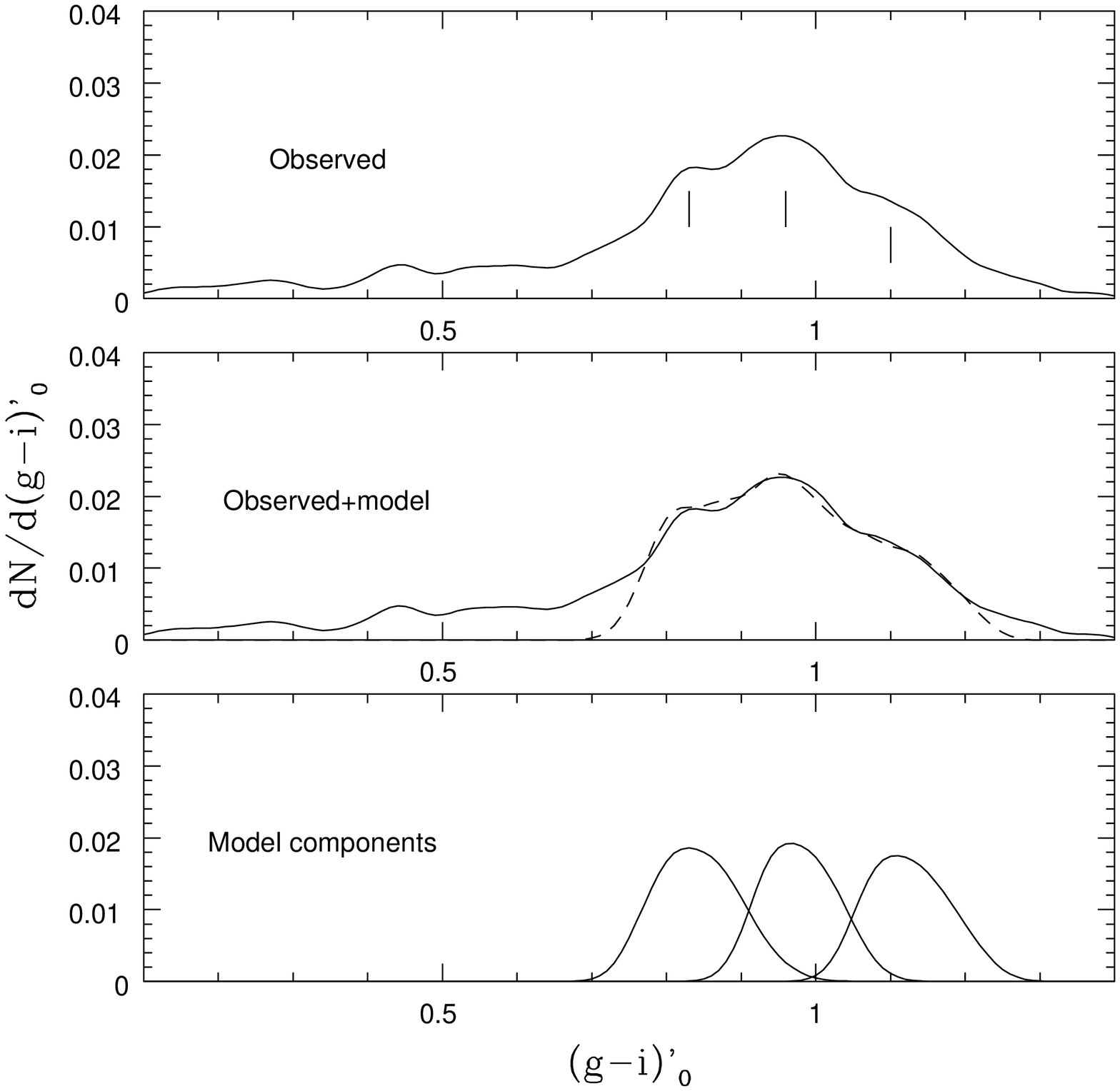}
    \caption{ Normalized $(g-i)'_{0}$ colour distribution for objects brighter than $g'_{0}=$25 and
     galactocentric radii from 90 to 270 arcsecs (upper panel). This distribution is compared with a three
     component models (dashed lines) in the middle panel. Each of these components are shown in the lower
     panel.}
    \label{fig:fig21}
\end{figure}

  An attempt to discriminate among the different GC populations on the basis of $(C-R)$ colours was
  presented in \citet{Richtler2012b}. In our case, instead of adopting  Gaussian colour distributions
  as these authors, we use an exponential dependence of the number of GCs as a function of $Z$
  (fractional mass chemical abundance of heavy elements)  \citep*[e.g.][]{Forte2007}, that we transform 
  to $(g-i)'_{0}$, which is our most sensitive index to metallicity, through a given model age-metallicity 
  relation.\\
  The procedure is similar to that described in \citet{Forte2013} and references therein. The 
  approach starts with a `seed' GC chemical abundance, $Z$, obtained from a Monte-Carlo generator controlled
  by a statistical distribution $N \approx$ $exp(-Z/Zs)$, within  a lowest ($Z_{i}$)  and maximum ($Z_{max}$)
  chemical abundances, and a scale parameter $Z_{s}$.\\
  The $(g-i)'_{0}$ colour distribution of the whole cluster sample shows an extended  blue tail that
  corresponds to presumably young clusters and field interlopers. We do not attempt to model these
  objects and restrict our fits to the blue, intermediate and red GCs. 
  For the `blue' GCs we adopted $Z_{i}=$ 0.02 $Z_{\odot}$  and $Z_{max}=$0.3 $Z_{\odot}$
  while both for the `intermediate' and `red' population we set $Z_{max}$=4 $Z_{\odot}$. The upper
  $Z$ cutoff adopted for the `blue' GCs seems appropriate for an adequate representation of the
  colour distributions of these clusters in a wide range of galaxy masses \citep[e.g.][]{Forte2014}.\\
  The $(g-i)'$ colour, determined from the colour-abundance relation, is then `blurred' by simulating
  observational errors as a function of the $g'$ magnitude. Each `seed' GC is also
  characterized by an apparent $g'$ magnitude generated adopting a  Gaussian GC luminosity function
  with a turnover at $g'$ $\approx$ 24.3 and a dispersion $\sigma_{g'}$=1.2 mag which is comparable
  to the values observed in giant ellipticals \citep[e.g.][]{Villegas2010}.\\
  We start with a tentative initial number of objects with a given chemical abundance $[Z/H]$ for each
  GC subpopulation, aiming  at reproducing the position of the modal colours and then iterate both these
  numbers and the corresponding $Z_{s}$  parameters until the output model gives the best match to the
  observed colour distribution. In this process, we ignored GC candidates bluer than $(g-i)'_{0}=$0.75 that
  possibly belong to a the `very blue' population.\\ 
  The quality criteria of the fit is to minimize the $rms$ of the observed
  vs. model numbers in the colour range $(g-i)'o=$ 0.75 to 1.35. More sophisticated approachs are probably
  not justified given the number of  involved parameters. An assessment of the consistency of the results 
  can be obtained by comparing the fits corresponding to two photometric samples defined through their
  limiting $g'_{0}$ magnitudes that we set at 23.50 and 25.0 mag.\\
  Table \ref{table_5} gives the set of parameters that provide the best representation of the `blue', `intermediate' 
  and `red' GC populations, shown in the middle and lower panels of Fig.~\ref{fig:fig13} for
  the GC sample brighter than $g'_{0}=23.5$. In this colour-magnitude domain, the `blue', `intermediate' and
  `red' GCs represent 87 percent of the sample. The remaining 13 percent are probably a combination of 
  young clusters and `blue' field interlopers.\\
  Increasing the limiting magnitude  to $g'_{0}=25.0$, and adopting a galactocentric range of 90 to 270
  arcsecs, leads to a GC sample 2.4 times larger.
  Model fit parameters for this sample are summarized in Table \ref{table_6},
  and their colour distributions, displayed in the middle and lower panels shown in Fig.~\ref{fig:fig21}.\\
  A comparison with the previous fit, shows an increase of the relative number of `intermediate'
  GCs (implying a steeper integrated luminosity function for these clusters) while the chemical abundance parameters
  show a small change that may be explained as a consequence of an increase of the field contamination
  and the larger photometric errors of the fainter objects.\\
\begin{table}
\centering
\caption{Model fit parameters for GCs brighter than g=23.5 and galactocentric radii from 45 to 270 arcsecs.}
\begin{tabular}{lccccccc}
\hline
\hline  
\textbf{Pop.}&\textbf{N}&\textbf{Z$_{s}$}&\textbf{Z$_{i}$}&\textbf{Z$_{max}$}&\textbf{Age}&\textbf{Z/Z$_{0}$}&\textbf{[Z/H]}\\
\hline
Blue&88&0.07&0.02&0.30&12&0.09&-1.11\\
Interm.&67&0.33&0.70&4&5&0.96&-0.03\\
Red&36&0.25&0.50&4&12&0.70&-0.16\\
\hline
Total&191&&&&&\\
\hline
\label{table_5}
\end{tabular}
\end{table}
\begin{table}
\centering
\caption{Model fit parameters for GCs brighter than g=25.0 and galactocentric radii from 90 to 270 arcsecs.}
\begin{tabular}{lccccccc}
\hline
\hline  
 \textbf{Pop.}&\textbf{N}&\textbf{Z$_{s}$}&\textbf{Z$_{i}$}&\textbf{Z$_{max}$}&\textbf{Age}&\textbf{Z/Z$_{0}$}&\textbf{[Z/H]}\\
\hline 
Blue&172&0.07&0.02&0.3&12&0.08&-1.17\\
Interm&168&0.40&0.60&4&5&0.88&-0.06\\
Red&113&0.35&0.50&4&12&0.69&-0.18\\
\hline 
Total&453\\
\hline 
\label{table_6}
\end{tabular}
\end{table} 
  The $[Z/H]$ abundances determined for the intermediate GCs in both model fits are slighlty sub-solar,
  a result that is in very good agreement with the spectroscopic solar abundance derived for three intermediate
  GCs  presented by \citet{Goudfrooij2001b}.\\
  The adoption of an age of 12 Gyr in modelling the intermediate GCs would lead to much lower chemical abundances of 
  $\approx$ 0.4 $Z_{\odot}$.\\
  The age we derive for intermediate GCs is larger than that by \citet{Richtler2012b} who
  find an age of 2 Gyr. Part of this discrepancy might arise in the different adopted models. On the other side
  their $(C-R)$ colours seem -0.13 mag. bluer than expected from the $(C-T_{1})_{0}$ vs $(g-i)_{0}$ relation (as
  show in Fig. \ref{fig:fig9}). A small  difference between the $(C-T_{1})_{0}$ and $(C-R)$ indices may be present
  \citep[e.g.][]{Geisler1996}, but we cannot explain the origin of the relatively large colour offset.\\ 
\section{Spatial distributions}
\label{sec5}
\subsection{Distribution on the sky}
\label{sec5.1}
\begin{figure}
	\includegraphics[width=\columnwidth]{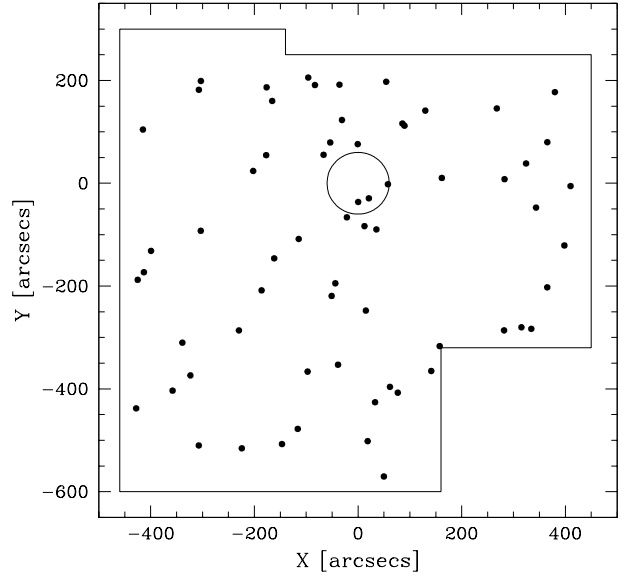}
    \caption{Distribution on the sky of 66 objects brighter than $g=23.50$ and
     $(g-i)'_{0}$ colours between 0.30 and 0.75. The reference circle, centered on
     the galaxy nucleus, has a 60 arcsecs radius.}
    \label{fig:fig22}
\end{figure}

\begin{figure}
	\includegraphics[width=\columnwidth]{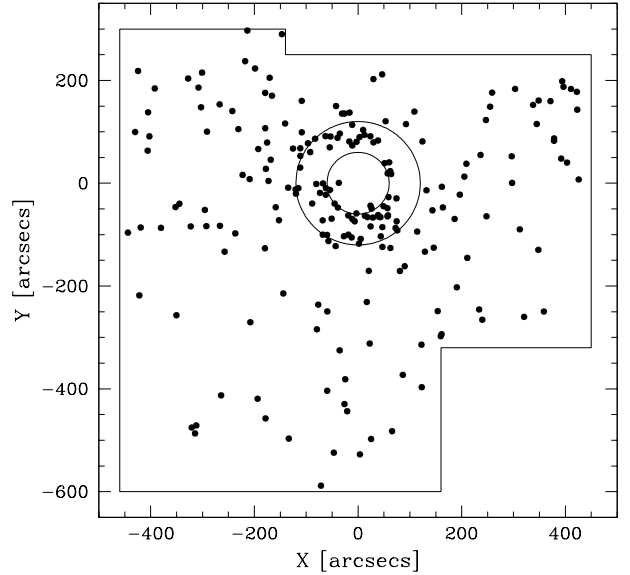}
    \caption{Distribution on the sky  of 229 objects with $g'_{0}$ from 23.5 to 25 and $(g-i)'_{0}$ from 
     0.30 to 0.75. The reference circles have 60 and 120 arcsecs radius.}
    \label{fig:fig23}
\end{figure}

\begin{figure}
	\includegraphics[width=\columnwidth]{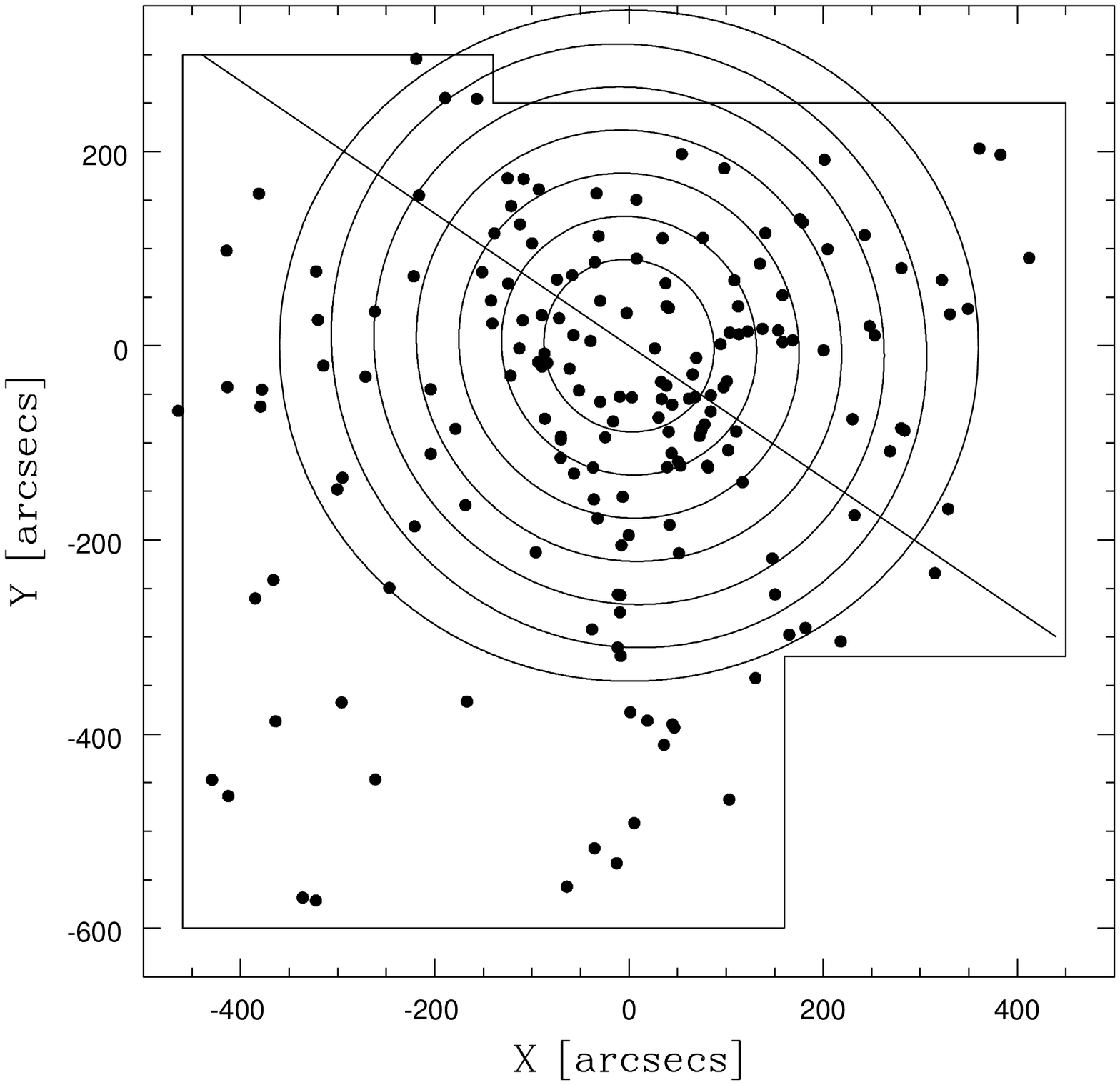}
    \caption{Distribution on the sky  of 177 objects brighter than $g=25.0$ and
    $(g-i)'_{0}$ colours between 0.75 and 0.90 (`blue' GCs). The ellipses (from $a=$90 to $a=$360 arcsecs)
    have a flattening $q=b/a=0.96$. The straight line indicates the position of
     the major axis of the galaxy.}
    \label{fig:fig24}
\end{figure}

\begin{figure}
	\includegraphics[width=\columnwidth]{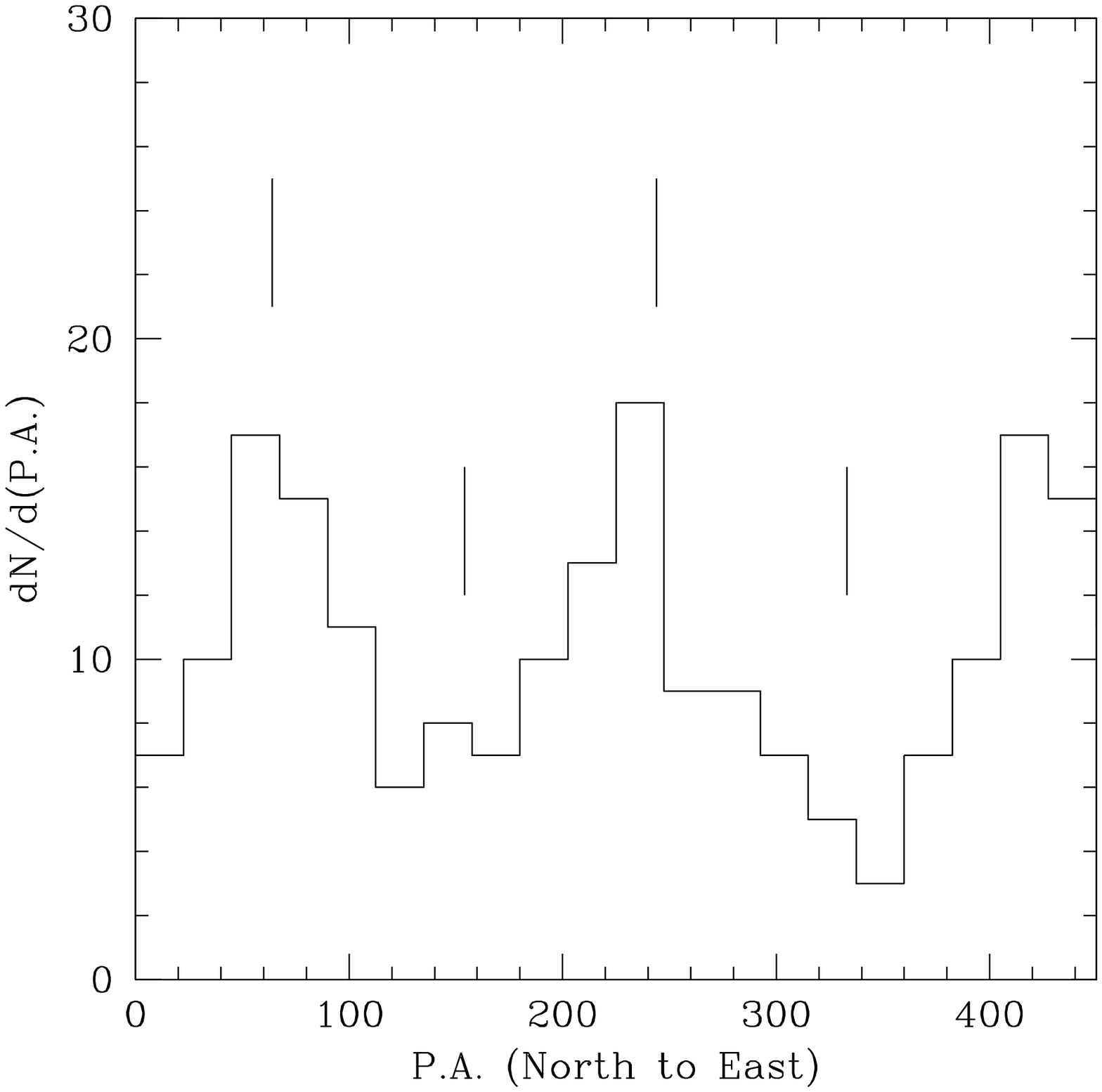}
    \caption{Azimuthal counts of `intermediate' GC candidates in a colour window defined between
 $(g-i)'_{0}$ 0.9 and 1.05. These counts correspond to a total of 155 clusters within a circular annulus
 with inner and outer radii of 90 and 220 arecsecs. The vertical lines
 show minimum and maximum values expected for an elliptical distribution with a $PA$ of 63 degrees.}
    \label{fig:fig25}
\end{figure}

\begin{figure}
	\includegraphics[width=\columnwidth]{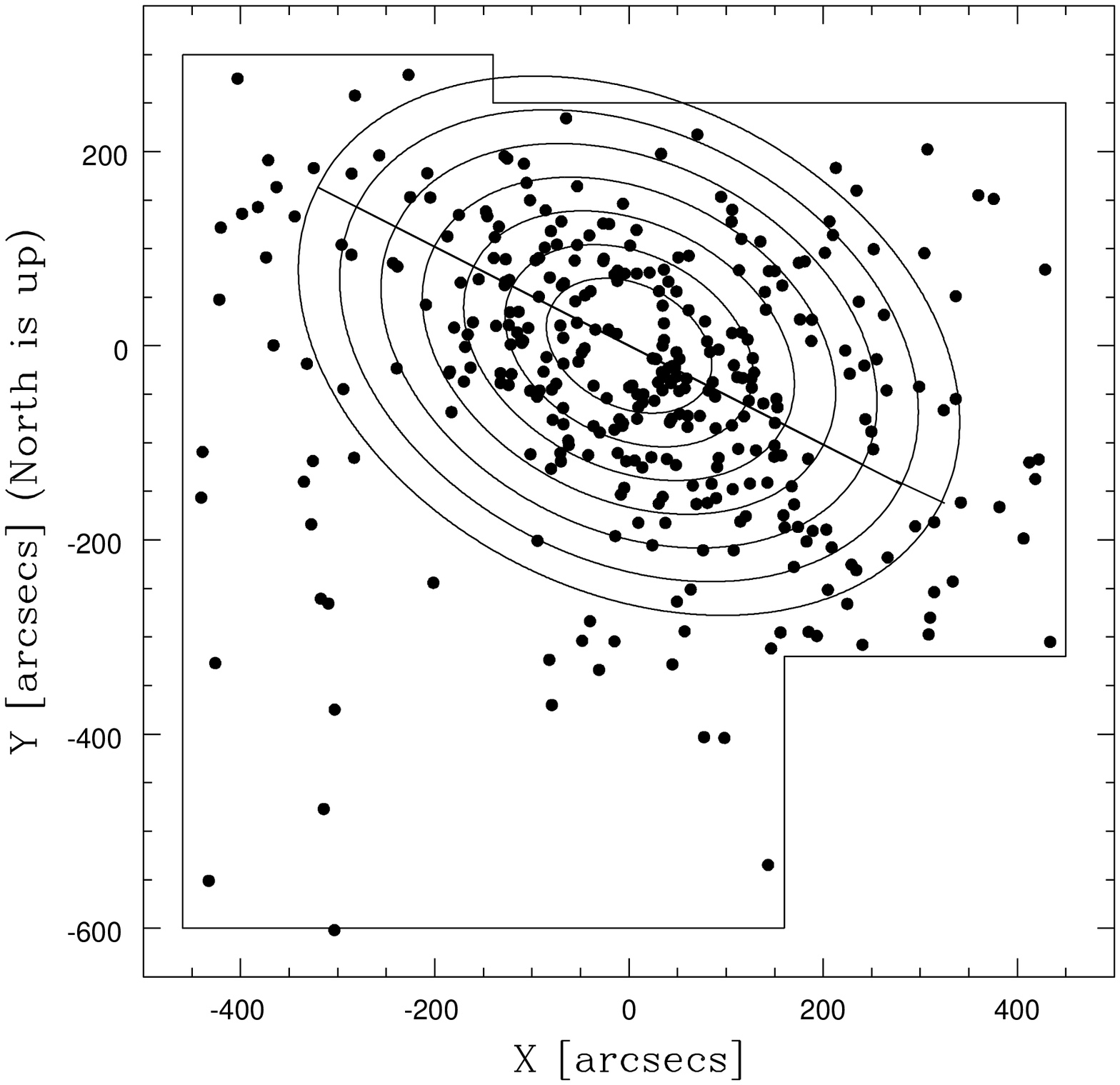}
    \caption{ Distribution on the sky of 340 objects brighter than $g'_{0}=25$ and $(g-i)'_{0}=0.90$ to
    1.05 (`intermediate' GCs). The ellipses (from $a=$90 to $a=$360 arcsecs) have a flattening $q=b/a=0.70$.
    The ellipses have a position angle $PA$ of 63 degrees.}
    \label{fig:fig26}
\end{figure}

\begin{figure}
	\includegraphics[width=\columnwidth]{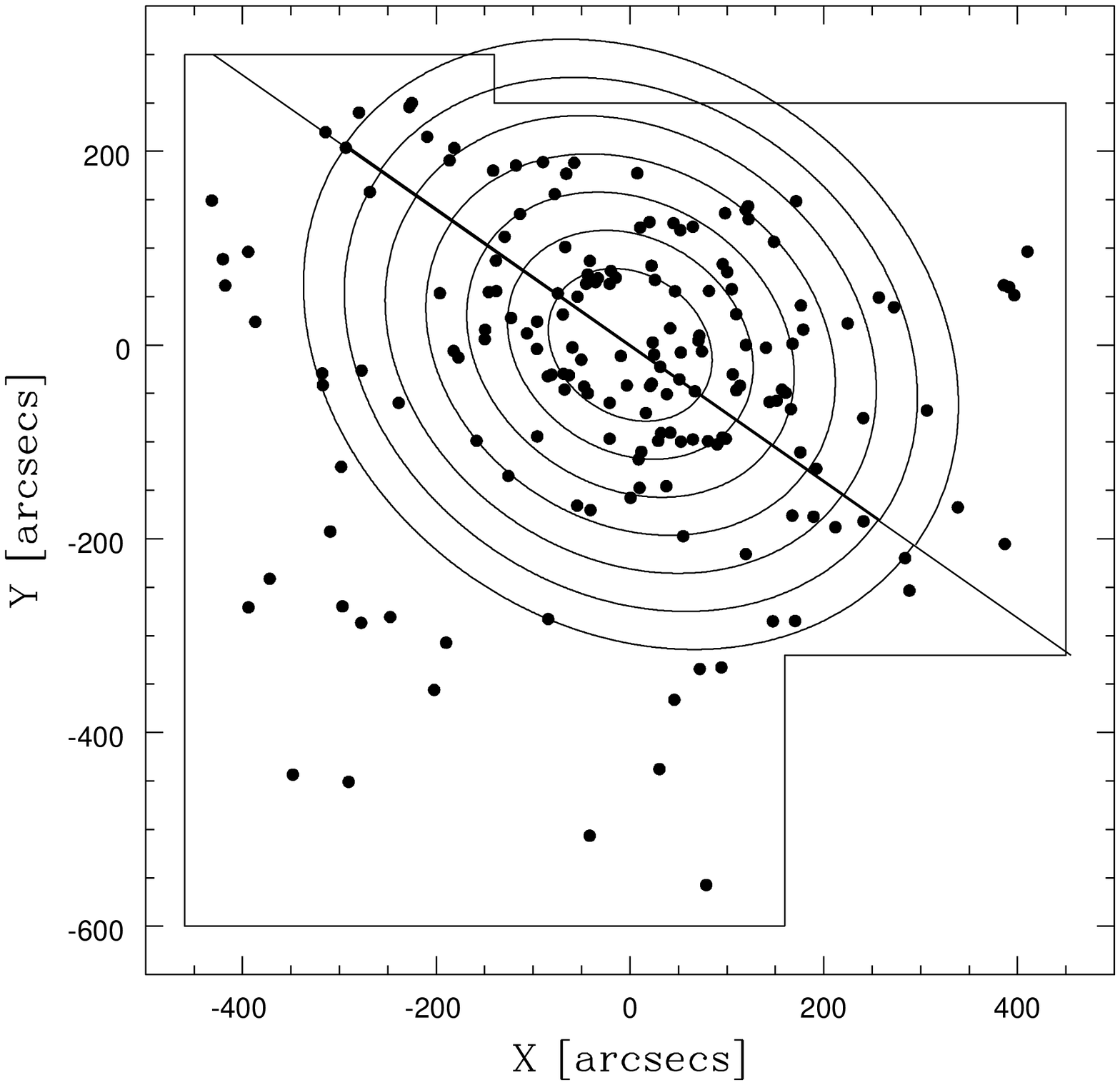}
    \caption{ Distribution on the sky of 170 red GC candidates brighter than $g'_{0}=25$ and 
    $(g-i)'_{0}=$ 1.05 to 1.35 (`red' GCs).
    The ellipses (from $a=$90 to $a=$360 arcsecs) have a flattening $q=b/a=0.81$. The straight line indicates the position of
     the major axis of the galaxy.}
    \label{fig:fig27}
\end{figure}

In this Section we analyze the projected areal
distribution of the different GC subpopulations on the sky. Previous analysis of this
subject have been presented in \citet{Gomez2001}, and in \citet{Richtler2012b}.\\
In order to isolate a given GC subpopulation, and decrease the eventual contamination 
from the colour-adjacent populations, we define colour windows using the results
of the model GC decomposition presented in previous sections. This leads to colour
ranges in $(g-i)'_{0}$ of 0.75 to 0.90 for the `blue' GCs, 0.95 to 1.05 for the
`intermediate' GCs and 1.05 to 1.35 for the `red' GCs. The position and widths of these
windows are shown in the lower panel of Fig.~\ref{fig:fig13}.\\
The distribution on the sky for 69 `very blue' objects brighter than $g'_{0}=$ 23.5 is depicted in Fig.~\ref{fig:fig22},
where these objects do not show a detectable concentration towards the centre of the galaxy. 
This might indicate that they are just field objects. However, as noted before, most of them 
 appear near the 1 Gyr model isochrone in the $(g-r)'_{0}$ vs $(r-i)'_{0}$ diagram.\\
On the other side, there are 229 objects within the same colour range, but fainter ($g'_{0}$ from 23.5 to 25), displayed
in Fig.~\ref{fig:fig23}. Among them, 62 appear closely packed in an annular region defined between 60 and 120 arcsecs 
in galactocentric radius. The areal density in this annulus is some 5 times larger than in the rest of the mosaic field 
suggesting that they are associated with $NGC~1316$.\\
\noindent `Blue' GC candidates, as displayed in Fig.~\ref{fig:fig24}, show a concentration
 towards the galaxy centre and follow a rather spheroidal distribution.\\
 In turn, Intermediate GCs exhibit a marked flattening and for this
 population we perform an analysis of their azimuthal distribution. Azimuthal counts within 
a circular annulus with a complete areal coverage (inner and outer radii of 90 and 220 arcsecs)
are displayed in Fig. \ref{fig:fig25} where vertical lines correspond to a position angle of 63
degrees. This result is in excellent agreement with a previous analysis by \citet{Gomez2001}.\\ 
The distribution of the intermediate GCs on the sky, as well as the elliptic annuli whose flattening and 
position angle were determined from the azimuthal counts, are displayed in Fig.~\ref{fig:fig26}.\\
Finally, Fig.~\ref{fig:fig27}, corresponds to the `red' GC candidates which exhibit a somewhat higher
flattening than that corresponding to the blue GCs.
\subsection{Areal density profiles}
\label{sec5.2}
\begin{figure}
	\includegraphics[width=\columnwidth]{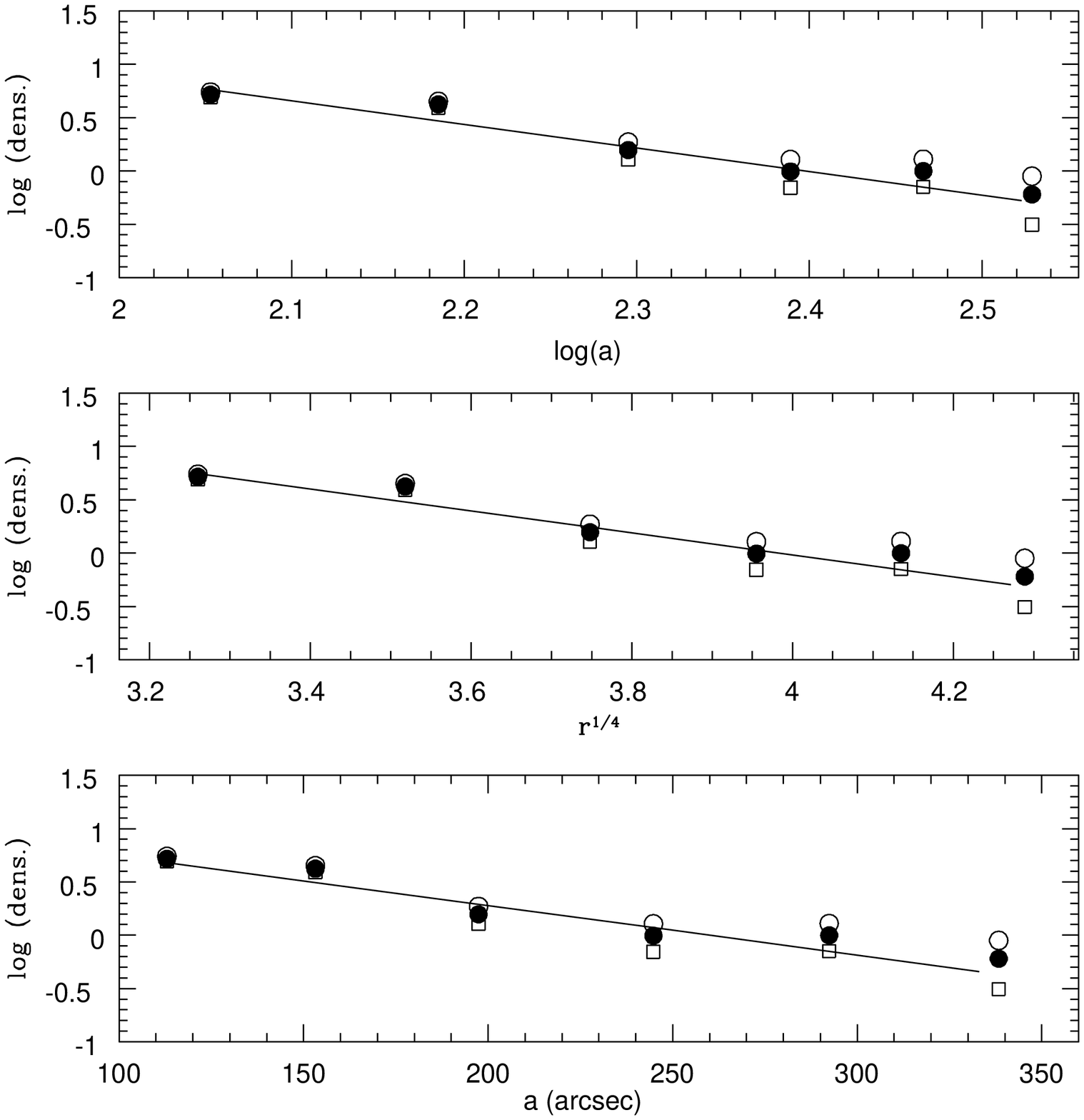}
    \caption{Projected density as a function of semimajor axis  for the `blue' clusters
    ($(g-i)'_{0}=$ 0.75 to 0.90 mag). A power law (upper diagram), $r^{1/4}$(middle), and
    disc-like (lower) profile fits are displayed.}
    \label{fig:fig28}
\end{figure}

\begin{figure}
	\includegraphics[width=\columnwidth]{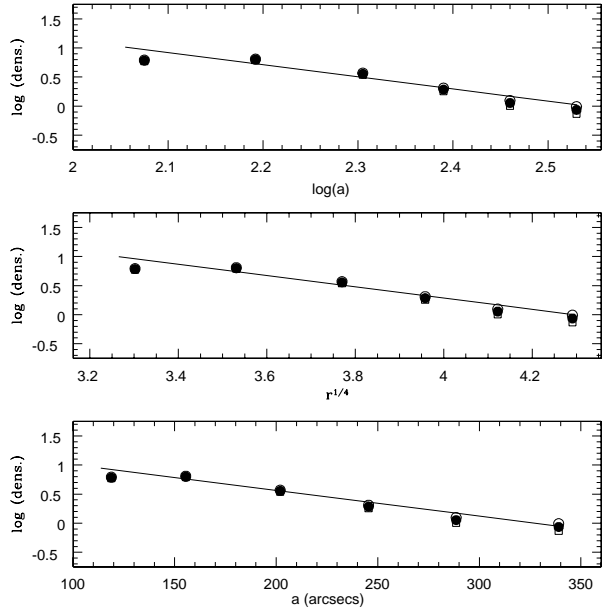}
    \caption{Areal density distribution for the `intermediate' clusters
    ($(g-i)'_{0}=$ 0.95 to 1.05 mag). A power law (upper diagram), $r^{1/4}$ (middle) and
    disc-like (lower) profile fits are displayed.}
    \label{fig:fig29}
\end{figure}

\begin{figure}
	\includegraphics[width=\columnwidth]{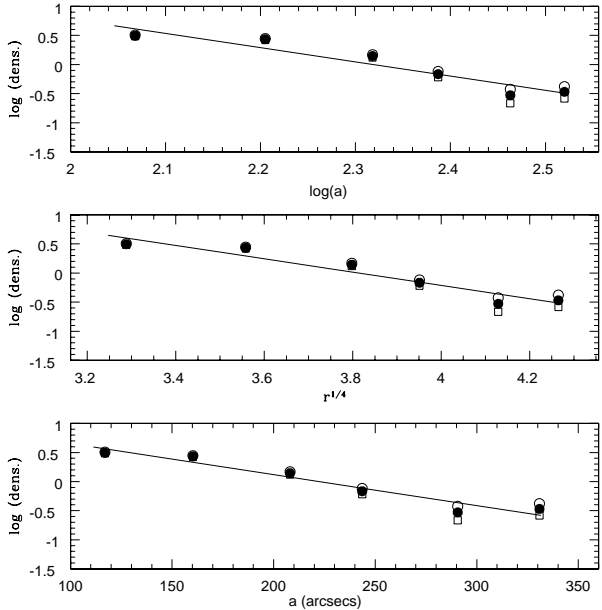}
    \caption{Areal density distribution for the `red' clusters
    ($(g-i)'_{0}=$ 1.10 to 1.35 mag). A power law (upper diagram), $r^{1/4}$ (middle) and
    disc-like (lower) profile fits are displayed.}
    \label{fig:fig30}
\end{figure}

 In this subsection we analyze the behaviour of the cluster areal distributions as a
 function of their corresponding semimajor axis adopting the colour windows described in the
 previous subsection.
 The areal densities correspond to a magnitude limit $g'_{0}=$ 25.0 and were derived in a region
 bounded by inner and outer semimajor axis of 90 and 360 arcsecs. Areal incompleteness of the
 outermost elliptical annuli (seen in Figs.~\ref{fig:fig24}, \ref{fig:fig26} and \ref{fig:fig27}) was
 taken into account when computing the projected densities.\\
 As we do not have a suitable field to estimate the level of the background contamination, we
 define a reference area including all objects with ordinates smaller than $y=$-300 arcsecs in
 Fig.~\ref{fig:fig6}. This region spans 50 arcmin$^{2}$. We then derive three density profiles for
 the different cluster populations. Each of these profiles assumes, respectively, null contamination (i.e., there are no
 background contamination; open dots); a `reference' background level assuming that half of the objects in the
 that area  are field interlopers (filled dots), and finally, that all of the objects in the 
 reference area are field objects (open squares). The resulting profiles then give an idea about
 the uncertainty connected with the adopted contamination density.\\
 As discussed before, the very blue clusters exhibit a very sparse distribution (except the concentration 
 detectable in the central regions of the galaxy) and we do not attempt the determination of a
 density profile for these objects.\\
 Figs.~\ref{fig:fig24}, \ref{fig:fig26} and \ref{fig:fig27} indicate that the blue, intermediate and red 
 GC subpopulations have  distinct spatial distributions and flattenings. An estimate of these flattenings was 
 obtained by computing the ratio of the second order momentums of the $y$ and $x$ coordinates defined in a 
 rotating frame centered on the galaxy nucleus. For this estimation, only objects within a circular annulus with a complete areal coverage were considered (inner and outer radii of 90 and 220 arcsecs).\\
 In the case of the blue and red GCs, with  low flattenings, the resulting position angles seem consistent
 with that of the major axis of the galaxy ($PA= 55 \deg$). In contrast, the intermediate GCs (as discussed before)
 exhibit a $PA$ 8 degrees larger than that.\\  
 Different scale laws have been adopted in the literature for the discussion of the GC density
 profiles. For example, power laws, de-Vaucouleurs like dependencies (i.e. $r^{1/4}$) or S\'ersic 
 profiles, a particular case of which ($n=1$) corresponds to disc-like structures.\\
 In our approach, we performed least squares fits adopting all these scaling laws for each of the
 GC subpopulations in an attempt to asses which one provides the best profile representation.\\ 
 The $lsq$ fits  given in Table \ref{table_7}, \ref{table_8} and \ref{table_9}
 correspond to  adopting the `reference background' and are shown in Fig.~\ref{fig:fig28}, Fig.~\ref{fig:fig29},
 and  Fig.~\ref{fig:fig30}. In these  diagrams, the open and filled circles, and the open squares belong
 to the three background level options described before.\\
 The uncertainties of the cluster counts within each annulus are comparable to the size of the plotting
 symbols.\\
\begin{table}
\centering
\caption{Projected density as a function of semimajor axis for the 115 `blue' GCs  with $(g-i)'_{0}=$0.75 to 0.90 mag ($q=0.96$,
 background=0.20 objects/arcmin$^{2}$).}
\begin{tabular}{lccc}
\hline
\hline  
\textbf{Scaling law}&\textbf{Slope}&\textbf{Zero point}&\textbf{rms}\\
\hline
Power&-2.132(0.156)&5.203(0.025)&0.062\\
$a^{1/4}$&-0.986(0.083)&4.022(0.029)&0.083\\
Disc&-0.0044(0.0006)&1.246(0.045)&0.111\\
\hline 
\label{table_7}
\end{tabular}
\end{table}
\begin{table}
\centering
\caption{Projected areal density fits for 210 `intermediate' GCs with $(g-i)'_{0}=$0.95 to 1.05 mag ($q=$0.70; background=0.12 objects/arcmin$^{2}$).}
\begin{tabular}{lccc}
\hline
\hline  
\textbf{Scaling law}&\textbf{Slope}&\textbf{Zero point}&\textbf{rms}\\
\hline
Power&-1.924(0.224)&4.75(0.036)&0.088\\
$a^{1/4}$&-0.892(0.094)&3.694(0.033)&0.080\\
Disc&-0.0040(0.0004)&1.188(0.030)&0.074\\
\hline 
\label{table_8}
\end{tabular}
\end{table}
\begin{table}
\centering
\caption{Projected areal density fits for 98 `red' GCs  with $(g-i)'_{0}=$1.05 to 1.35 mag ($q=$0.81; background=0.08
objects/arcmin$^{2}$.)}
\begin{tabular}{lccc}
\hline
\hline  
\textbf{Scaling law}&\textbf{Slope}&\textbf{Zero point}&\textbf{rms}\\
\hline
Power&-2.328(0.417)&5.409(0.066)&0.164\\
$a^{1/4}$&-1.082(0.189)&4.140(0.066)&0.161\\
Disc&-0.0049(0.0009)&1.107(0.067)&0.166\\
\hline 
\label{table_9}
\end{tabular}
\end{table}
\begin{figure}
	\includegraphics[width=\columnwidth]{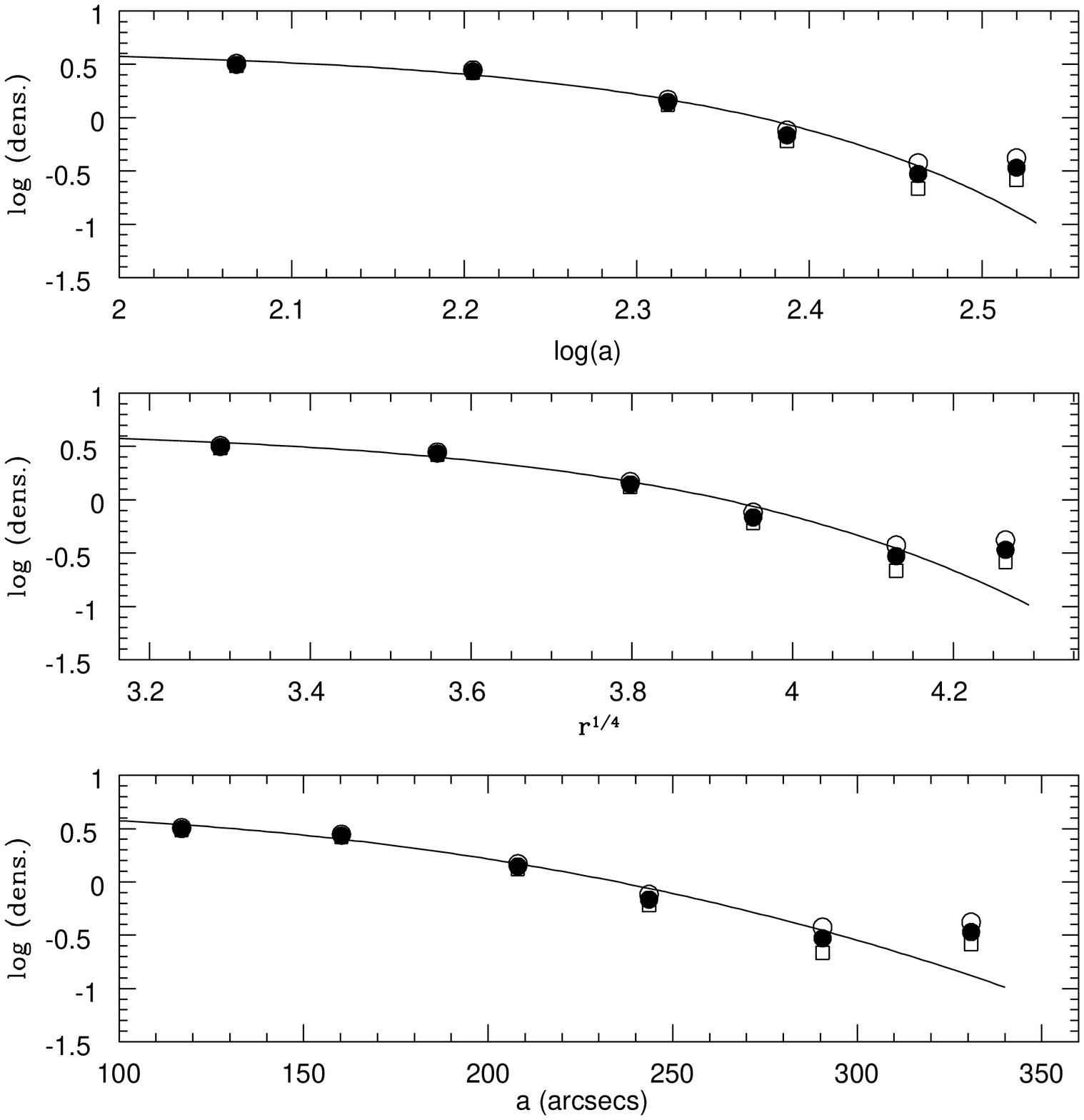}
    \caption{Areal density distribution for the `red' clusters($(g-i)'_{0}=$ 1.10 to 1.35 mag).
    A S\'ersic profile with a scale lenghth of 200 arcsecs and $n=$ 0.4 (see text) is shown as solid lines.}
    \label{fig:fig31}
\end{figure}
 Tables \ref{table_7} and \ref{table_8} show that, in terms of the rms of the fits, a power law gives the best
 representation of the areal density behaviour for the `blue' clusters, while 
 `intermediate' GCs are better represented by a disc ($n=1$). These two populations
 (despite their different flattenings and semimajor axis position angle) exhibit rather
 similar slopes along their respective semimajor axis $a$.\\
 In the case of the `red' GCs, and as shown in Table \ref{table_9}, none of the scaling 
 laws seem to provide an acceptable representation since the fits leave large and systematic 
 residuals. For these clusters, and as an alternative, we show  a S\'ersic profile, with an
 index $n=$0.4 and a scale parameter of 200 arcsecs, that provides a better representation 
 of the density profile, as depicted in Fig.~\ref{fig:fig31}.\\
\section {Radial Velocities}
\label{sec6}
\begin{figure}
	\includegraphics[width=\columnwidth]{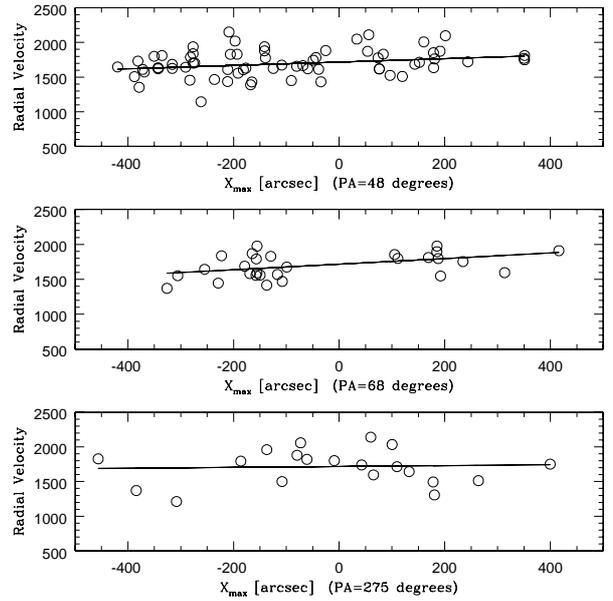}
    \caption{Radial velocities from \citet{Richtler2014} as a function of the axis that
     exhibits the maximum velocity gradient for 67 blue GCs (upper panel), 28 GCs intermediate (middle panel)
     and 20 red GCs (lower panel). $X_{max}=$0 indicates the center of the galaxy.
The typical RV errors ($\sim$50 $km/sec$) are smaller than the size of the plotting symbols.}

\label{fig:fig32}
\end{figure}

 A revision of our photometry shows that 124 objects have radial velocties (RV) measured by \citet{Richtler2014}.
 Among them, 115  fall within the colour windows we adopted to isolate the blue (67 objects), intermediate (28 objects) 
 and red GC candidates (20 objects). The nine objects excluded are slightly bluer than the window we use to define the blue GCs, but in any case, their inclusion in the figure of the blue CG has no effect on the analysis.\\
 For each of these groups we found the position angle on the sky of the axis that maximizes the radial velocity
 gradients ($X_{max}$), as shown in Fig.~\ref{fig:fig32}. The position angle is measured from North towards East until reaching
 the section of the axis that contains the objects with receeding radial velocties (i.e., velocities larger than
 the GC mean radial velocity).\\
 The parameters of these gradients are listed in Table \ref{table_10}
 for each GC subpopulation. This table indicates that both the blue and intermediate GCs exhibit significant and
 different RV gradients while red GCs do not show a detectable gradient and display the largest $rms$ of the velocity
 residuals.\\
\begin{table}
\centering
\caption{Maximum radial velocity gradients for the blue, intermediate and red GC populations.}
\begin{tabular}{lcccc}
\hline
\hline 
\textbf{Pop.}&\textbf{N$_{GCs}$}&\textbf{Slope(Err)}&\textbf{PA}&\textbf{residual rms}\\ 
&&&degrees&$km/sec$\\
\hline
Blue&67&0.22($\pm$ 0.11)&48&195\\
Interm.&28&0.42($\pm$ 0.08)&68&154\\
Red&20&0.06($\pm$ 0.28)&275&260\\
\hline
\label{table_10}
\end{tabular}
\end{table}
  \citet{Richtler2014},  noticed that the kinematic axis of the stellar halo in $NGC~1316$, $PA=$ 71 degrees,
 is missaligned with the optical axis of the galaxy. Besides, \citet{McNeil2012} also detect such a
 deviation, on the basis of the analysis of the radial velocities of 490 planetary nebulae, and find $PA=$64 degrees.\\
 The $PA$ of the maximum radial velocity gradient of the intermediate GCs, falls between those last
 angles and suggests that, at least these clusters, have a kinematic link with field stars.\\
\section{GC Luminosity functions}
\label{sec7}
\begin{figure}
	\includegraphics[width=\columnwidth]{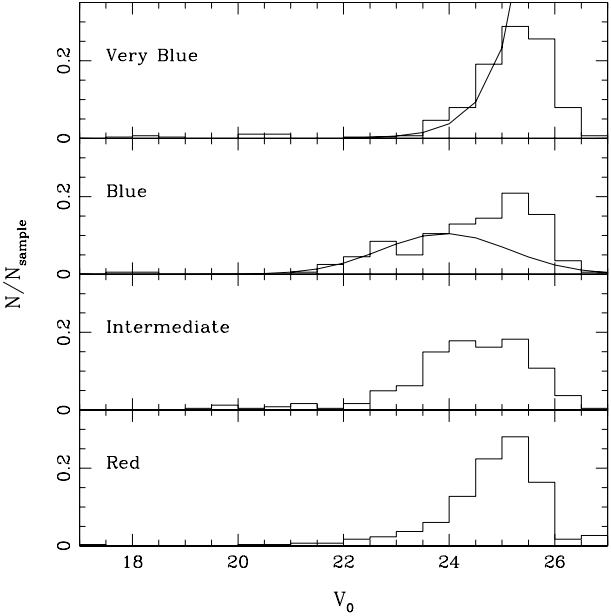}
    \caption{Integrated luminosity function for the `very blue' clusters (upper pannel), `blue',
    `intermediate', and `red' (lower pannel) globular clusters. As a reference, a Gaussian with a turnover at $V_{0}=23.93$ and dispersion $\sigma=$ 1.2 mag.,
    is plotted for the `blue' globulars. An exponential is shown for the very blue objects.}
    \label{fig:fig33}
\end{figure}

 A thorough discussion of the characteristics of the integrated luminosity function of the cluster 
 candidates  requires a deep comparison field which we do not have. However, and adopting the same colour
 `windows' defined in Section \ref{sec5}, we present a first analysis of the apparent magnitude 
 distribution of the four cluster subpopulations that is meaningful for objects brighter than 
 $g'_{0}$=25 and with galactocentric radius ranges of 90 to 270 arcsecs, for which our completeness level is close to 90 percent.\\
 With the aim of comparing this analysis with previous work, we transformed the GC $g'_{0}$ magnitudes
 to $V_{0}$ magnitudes through the relation given by \citet{Chonis2008}: $V=g'-0.011-0.587 \times (g-r)'$.\\
 The resulting luminosity functions, uncorrected for completeness and normalized by total number of each population ($N_{sample}$), are displayed in the four panels of Fig.~\ref{fig:fig33} (top to bottom: `very blue', `blue', `intermediate' and `red' clusters).\\
 The `very blue' objects show a steep exponential behaviour, starting at $V$ $\approx$ 22.0 and with a scale parameter
 of 0.55 mag.\\ 
 `Blue' GCs seem well represented by a reference Gaussian curve with a turnover at $Vo=23.93$ which we estimate adopting
 the distance modulus given by \citet{Cantiello2013} and assuming $M_{V}(TO)=-7.66$ (\citealt{Dicriscienzo2006}, 
 obtained from the low metallicity halo   clusters in the MW and M31), and a dispersion $\sigma=1.2$, typical for GCs
 in giant ellipticals.\\
 The increase in the number of objects above the Gaussian, for objects fainter than $V_{0}=24$ is possibly explained
 by the presence of field interlopers (see Fig.~\ref{fig:fig10}).\\
 The luminosity function of the `intermediate' GCs is significantly broader than those of the remaining GC families and 
 shows a significant fraction of GC candidates brighter than $V_{0}=22$, reaching $\approx$ 19 which, as discussed by
\citet{Goudfrooij2001b}, would be compatible with an age younger than that of the `blue' GCs. For this population
 \citet{Gomez2001} determine a turnover magnitude $V$=24.82.\\
 Finally, `red' GCs, also display a rapid increase in number. This behaviour of the red GCs was previously noticed by
 \citet{Goudfrooij2004}. The presence of a hypothetical turnover some 0.5 mag fainter than that of the `blue' GCs that
 could be justified by metallicity effects (\citealt*{Ashman1995}), cannot be ruled out.\\
\section{Conclusions}
\label{sec8}
  The discussion of the  $gri'$ photometry of cluster candidates in $NGC~1316$ presented in
  this paper, confirms earlier results that pointed out the complexity of the cluster
  system in this galaxy.\\
  We identify four cluster subpopulations with distinct characteristics, namely:\\
 
 a) Very blue clusters.
  \noindent These cluster candidates occupy a $(g-i)'_{0}$ colour range from 0.30 to 0.75. On one side, objects
  brighter than $g'_{0}=$23.5, show a very sparce distribution. Most of them show a colour distribution 
  compatible with  a 1 Gyr isochrone. Given the lack
  of unicity in the colour-abundance relation for that age, the situation remains largely undetermined and
  requires a spectroscopic analysis to clarify the nature of these objects.\\
  GC number $119$ in \citet{Goudfrooij2001b}, falls in that colour range and is a confirmed member of the
  $NGC~1316$ system, on the basis of its radial velocity. This cluster has been identified by \citet{Richtler2012b} 
  as a presumably very young cluster.\\
  These bright cluster candidates are an intriguing subpopulation that, if confirmed as such, would indicate
  the existence of an intense burst of cluster formation widespread on the whole body of the galaxy.\\
  In turn, fainter candidates within the same colour range, show a marked increase of the areal density
  in an annular region (60 to 120 arcsecs) around the center of the galaxy. The relatively large errors
  of the photometry for these faint clusters, prevent a significant analysis of their position in the two colour
  diagram. Their exponential luminosity function suggests that they might rather be massive young
  clusters than GCs.\\
  
  b) Blue GCs.
  \noindent This subpopulation exhibits a colour peak at $(g-i)'_{0}$=0.82, similar to those observed for the blue
  GCs in elliptical galaxies. As noticed by \citet{Goudfrooij2001b} their photometric features are compatible
  with the old low metallicity GCs in the MW halo.\\

 c) Intermediate GCs.
  \noindent The colours of these clusters correspond to a 5 Gyr old population in the frame of the Bressan et al.
  models. This age, and the mean chemical abundance we derive  for these clusters (slighthly subsolar),
  are comparable with the spectroscopic age (3 $\pm$ 1 Gyr) and abundance (Solar) obtained by \citet{Goudfrooij2012} 
  for three GCs which, according to their colours, belong to this population.\\
  The spatial distribution of the intermediate GCs exibits a flattenig $q$ $\approx$  0.7 and a position 
  angle of $63$ degrees, i.e., some 8 degrees larger than that of the semimajor axis of the $NGC~1316$ halo.
  \citet{Gomez2001} find  the same position angle for what they call `red' GCs. Their areal density profile can be  fit
  by S\'ersic profile with $n=$ 1 (disc-like). \\
  An analysis of the radial velocities  of 28 intermediate GCs (with data from \citealt{Richtler2014}) shows marked
  similitude with the kinematics of the stellar halo of the galaxy. This result argues in favor of a connection
  between the intermediate GCs and field stars.  The presence of an intermediate age stellar population was 
  pointed out by \citet{Cantiello2013} through a comparison of their SBF colours and SSP model colours.\\

 d) Red GCs.
  \noindent The colours of this population are compatible with those of old red GCs in elliptical galaxies.
  Their spatial distribution, spheroidal and slightly flattened, is coherent with a bulge-like high
  chemical abundance population (somewhat less than that of the intermediate GCs). They seem clearly
  distinct from the intermediate clusters. If the red GCs were coeval with the intermediate clusters,
  their chemical abundance should be three to four times larger than those we find on the basis of the
  model fits, not a very likely situation.\\

  A summary of all these features then indicates the existence of a rather spherical low metallicity halo,
  and of a more chemically enriched  and flattened bulge, coexisting with an `intermediate age' flattened 
  spheroid (or even a thick disc?) which exhibits photometric and kinematic similarities with the galaxy
  halo. This scenario is compatible with \citet{McNeil2012} who suggest that $NGC~1316$ may represent the early
  stages of a system that would evolve to become a `Sombrero' like galaxy ($NGC~4594$) through a series of mergers.
  The so called `very blue' clusters, for which we find a tentative age of 1 Gyr,  may be the tracers of the 
  last of these events.\\
   
\section*{Acknowledgements}

This work was funded with grants from Consejo Nacional de Investigaciones
Cientificas y Tecnicas de la Republica Argentina, and Universidad Nacional
de La Plata (Argentina). Based on observations obtained at the Gemini Observatory, 
which is operated by the Association of Universities for Research in Astronomy, Inc., under a cooperative agreement with the NSF on behalf of the Gemini partnership: the National 
Science Foundation (United States), the National Research Council (Canada), 
CONICYT (Chile), the Australian Research Council (Australia), Minist\'{e}rio da 
Ci\^{e}ncia, Tecnologia e Inova\c{c}\~{a}o (Brazil) and Ministerio de Ciencia, 
Tecnolog\'{i}a e Innovaci\'{o}n Productiva (Argentina). 
The Gemini program ID are GS-2008B-Q-54 and GS-2009B-Q-65. This research has made use of the 
NED, which is operated by the Jet Propulsion Laboratory, Caltech, under contract with the 
National Aeronautics and Space Administration.\\
We thank the referee for her/his significant contributions which greatly improved this work.\\

\bibliographystyle{mnras}
\bibliography{Sesto_etal.bib} 
\bsp	
\label{lastpage}
\end{document}